\documentclass[journal,10pt]{IEEEtran}
\usepackage{amsthm}
\usepackage{amssymb}
\usepackage{amsmath}
\usepackage{epsfig}
\usepackage{epstopdf}
\usepackage{cite}
\usepackage{bm}
\usepackage[noend]{algpseudocode}
\usepackage{lettrine}
\usepackage{algorithmicx,algorithm}
\IEEEoverridecommandlockouts

\graphicspath{ {figures/}}

\begin{document}
	
	\title{Symbol-Level Precoding for MU-MIMO System with RIRC Receiver}

	\author{Xiao Tong, Ang Li,~\IEEEmembership{Senior~Member,~IEEE}, Lei Lei,~\IEEEmembership{Member,~IEEE}, Fan Liu,~\IEEEmembership{Member,~IEEE}, \\ Fuwang Dong,~\IEEEmembership{Member,~IEEE}
	}
	
	\markboth{}
	{}
	
	\maketitle
	
	\begin{abstract}
		Consider a multiuser multiple-input multiple-output (MU-MIMO) downlink system in which the base station (BS) sends multiple data streams to multi-antenna users via symbol-level precoding (SLP), where the optimization of receive combining matrix becomes crucial, unlike in the single-antenna user scenario. We begin by introducing a joint optimization problem on the symbol-level transmit precoder and receive combiner. The problem is solved using the alternating optimization (AO) method, and the optimal solution structures for transmit precoding and receive combining matrices are derived by using Lagrangian and Karush-Kuhn-Tucker (KKT) conditions, based on which, the original problem is transformed into an equivalent quadratic programming problem, enabling more efficient solutions. To address the challenge that the above joint design is difficult to implement, we propose a more practical scheme where the receive combining optimization is replaced by the interference rejection combiner (IRC), which is however difficult to directly use because of the rank-one transmit precoding matrix. Therefore, we introduce a new regularized IRC (RIRC) receiver to circumvent the above issue. Numerical results demonstrate that the practical SLP-RIRC method enjoys only a slight communication performance loss compared to the joint transmit precoding and receive combining design, both offering substantial performance gains over the conventional BD-based approaches. 
	\end{abstract}
	
	
	\IEEEpeerreviewmaketitle
	
	\section{Introduction}\label{introduction}
	\lettrine[lines=2]{M}{ultiple}-input multiple-output (MIMO) has emerged as a revolutionary communication technique over the past decades. It enables higher data throughput and improved signal quality without the need for increased transmission power or bandwidth. By simultaneously multiplexing multiple data streams to multiuser (MU) in a spatial manner, MIMO technology has received significant attention from both the academia and industry, offering immense potential for the development of next-generation wireless communication systems \cite{1,2,3}. While MIMO technology offers numerous advantages and additional degrees of freedom, it also exhibits certain limitations when implementing MU-MIMO systems. The utilization of multiple antennas for transmission and reception introduces various challenges. These include signal pollution arising from multi-channel interference and noise \cite{M1,M2}. Additionally, the decoding efficiency is hampered by the inability of multiple users to decode signals jointly\cite{M3,M4}. To address these issues, precoding has been widely recognized as an effective strategy in MU-MIMO systems\cite{4,5,6,7,8,252,26,27,9,10,11,12,131,13,14,15,16,17,18,21,19,20,22}.
	
	The utilization of precoding techniques leverages the channel characteristics for transmit signal design to achieve enhanced communication performance. When the base station (BS) possesses knowledge of the channel state information (CSI), advanced nonlinear precoding methods such as dirty-paper coding (DPC)\cite{4}, Tomlinson-Harashima precoding (THP)\cite{5}, and vector perturbation (VP)\cite{6} can be employed to improve system performance. However, the complexity associated with these nonlinear methods hinders their practical implementation. To address this challenge, low-complexity linear precoding schemes have been proposed\cite{7,8,25,26,27}, including zero-forcing (ZF)\cite{7}, regularized zero-forcing (RZF)\cite{8}, block diagonalization (BD) \cite{25}, regularized block diagonalization (RBD) \cite{26}, generalized minimum mean-squared error (MMSE) \cite{27}.
	
	Additionally, downlink precoding schemes based on optimization have gained significant attention\cite{9,10,11,12}. These schemes aim to maximize the minimum user's signal-to-interference-plus-noise ratio (SINR) under a total transmit power budget\cite{9,10} or a per-antenna power budget\cite{11}. However, the complexity and coupled nature of the SINR optimization problem make it difficult to obtain closed-form solutions. To decouple the problem and enable an analytical closed-form solution, a leakage-based precoding scheme has been proposed in \cite{12}, which maximizes the signal-to-leakage-and-noise ratio (SLNR) simultaneously for all users, where closed-form solutions can be obtained. Nonetheless, the aforementioned precoding schemes primarily focus on interference suppression to ensure communication performance, disregarding the fact that interference can be beneficial and further exploited at the symbol level\cite{131,13,14,15,16,17,18}. 
	
	\subsection{Previous Work on Symbol-Level Precoding}\label{PW}
	Symbol-level precoding (SLP) is an approach that capitalizes on the information of both CSI and modulated data symbols to design the precoding strategy on a symbol-by-symbol basis\cite{131,13,14}, which means that both the amplitude and phase of the interference signals can be controlled judiciously to become constructive\cite{15,16,17,18}. This allows the interference to act as an additional source of the desired signal, thereby improving communication performance. The SLP scheme for multiuser multiple-input single-output (MU-MISO) downlink systems, which addresses transmit power minimization and SINR balancing for phase shift keying (PSK) modulated signals, was initially proposed in\cite{21}. Simulation and analysis demonstrate that the SLP scheme outperforms the conventional block-level precoding (BLP) methods. To mitigate computational costs, the work in \cite{19} further derives the optimal precoding structure and formulates the problem as a simplex quadratic programming (QP) problem. An iterative closed-form scheme is then proposed to obtain the optimal precoding matrix. Additionally, the authors extend this work to scenarios involving multi-level modulations in \cite{20}, where they consider the relationship between the number of served users and transmit antennas. In both cases where the number of served users is either more or less than the number of transmit antennas, the precoding problems are transformed into pre-scaling operations using QP optimization techniques.
	
	Indeed, while SLP can provide benefits in terms of improved communication performance, it does come with the tradeoff of symbol-by-symbol complicated signal processing operations. The aforementioned schemes involve high computational complexity, this is due to the need for iterative optimization procedures, solving quadratic programming problems, and performing symbol-level calculations for each transmission symbol. As a result, the computational costs associated with these approaches can be demanding, limiting their practical applicability in real-time systems or scenarios with resource-constrained devices. Therefore, a more practical approach is to strike a balance between communication performance and computational cost. The work in \cite{22} proposes a constructive interference-based block-level precoding (CI-BLP) approach for MU-MISO downlink systems. In this approach, a constant precoding matrix is applied to a block of symbol slots within a channel coherence interval, significantly reducing the total number of optimization problems that need to be solved. Furthermore, \cite{23} introduces a low-complexity grouped SLP (G-SLP) approach. In this method, all users are divided into several groups, allowing for the exploitation of intra-group interference through SLP while suppressing inter-group interference to reduce computational complexity. Furthermore, efficient G-SLP design algorithms are developed to solve the man-min fairness problem and power minimization problem.
	
	It is worth noting that the majority of existing works on SLP schemes primarily concentrate on the MU-MISO downlink systems, while only specific studies have addressed the challenges of MU-MIMO SLP problems and considered the design aspects of both transmit precoding and receive combining matrices \cite{31,32}. The work presented in \cite{31} introduces a joint design of symbol-level precoding (SLP) and receive combining (SLP-RC) for MU-MIMO systems. The objective of this approach is to minimize total transmit power while maintaining a satisfactory symbol error rate. To efficiently address this problem, a successive convex approximation algorithm is proposed. Furthermore, in \cite{32}, the investigation is extended to SLP-RC design with the additional complexity of one-bit transmission constraints. To tackle this NP-hard problem, a successive upper-bound minimization (SUM) and alternating direction method of multipliers (ADMM)-based algorithm is proposed. It is important to note that the SLP approaches discussed in \cite{31,32} for MU-MIMO systems require knowledge of the transmit signals for the design of the receive combining matrix. However, this requirement is highly impractical. As a result, finding more practical and efficient methods for designing the receive combining matrix in MU-MIMO systems using the SLP method remains an active and ongoing area of research. 
	
	\subsection{Contributions}\label{Contri}
	This paper studies the design of practical symbol-level transmit precoding and receive combining designs for MU-MIMO communication systems, where closed-form precoding and combining structures are derived. The focus is to exploit CI for generic PSK modulations to maximize the distance between the constructive region and the detection thresholds, which guarantees the communication performance with the best possible effort, and two specific SLP-RC designs are introduced. Initially, the transmit precoding matrix and receive combining matrix are jointly optimized to maximize communication performance subject to constraints of the total transmit power budgets, and the non-strict CI constraints. To solve the coupled optimization problem, an alternating optimization (AO) algorithm is utilized, which transforms the problem into two sub-problems, where the transmit precoding matrix and the receive combining matrix are updated iteratively. By analyzing the Lagrangian and Karush–Kuhn–Tucker (KKT) conditions of the two sub-problems, the paper derives the closed-form structures of the optimal transmit precoding matrix and receive combining matrix. This derivation leads to an equivalent simpler optimization problem, and the convergence can be guaranteed. To overcome the dependence of the receive combining matrix on the data information, the paper proposes to adopt the mature IRC receiver as the receive combiner for the users. By proving that the transmit SLP matrix is rank one, which makes the conventional IRC receiver not directly applicable, a regularized IRC (RIRC) is further proposed, finally leading to a practical SLP design for MU-MIMO system where the receive combiner is irrelevant to the transmit signals.
	
	For reasons of clarity, we summarize the contributions of this paper as:
	\begin{enumerate}
		\item This paper proposes a symbol-level joint transmit precoding and receive combining scheme to maximize the communication performance for MU-MIMO system, where the AO algorithm is used to solve the joint optimization problem.
		\item By analyzing the Lagrangian and KKT conditions, the closed-form solutions and more efficient algorithms are obtained for the proposed optimization problems. Furthermore, the convergence and computational complexity of the proposed algorithm are analyzed.
		\item The dependence of the receive combiner on the data information is addressed by introducing the RIRC receiver, which circumvents the issue that conventional IRC receiver is not applicable due to the rank-one nature of the transmit SLP matrix.
	\end{enumerate}
	
	Simulation results demonstrate that the practical SLP scheme with RIRC receiver experiences only a minor loss in communication performance compared to the joint transmit precoding and receive combining design with reduced complexity, both of which exhibit significant performance gains over conventional BD-based approaches.
	
	\subsection{Organization and Notations}\label{OAN}
	The rest of this paper is organized as follows. Section \ref{SMPF} introduces the considered MU-MIMO system model, then a brief review of SLP is given and the corresponding formulated problem is introduced. Section \ref{JSLTARBD} includes the analysis for joint symbol-level transmit precoding and receive combining optimization problem and the closed-form solution structures are derived. The problem is extended to a more practical scenario where the IRC receiver is introduced and the RIRC receiver is proposed in Section \ref{SLPWIR}. In Section \ref{CCCA}, the convergence and computational complexity are analyzed for proposed algorithms. The simulation results are shown in Section \ref{Simulation}, and Section \ref{Conclusion} concludes the paper.
	
	Notations: a, $ \bf a $ and $ \bf A $ denote scalar, vector, and matrix, respectively. $\textbf{I}$ is the identity matrix. $\mathbb{C}^{M\times N}$ ($\mathbb{C}^{M \times 1}$) represents a complex-valued  $ M \times N $ matrix ($ M \times 1 $ vector). The transpose, conjugate and complex conjugate transpose of a matrix or vector are denoted by using $(\cdot)^T$, $(\cdot)^*$ and $(\cdot)^H$. The inverse and trace of a matrix are $(\cdot)^{-1}$ and $\text{Tr}(\cdot)$. $|\cdot|$ denotes the absolute value of a real number or the modulus of a complex number, and $\Vert \cdot \Vert_2$ represents the ${l_2}$-norm. $ \mathcal{R}(\cdot) $ and $ \mathcal{J}(\cdot) $ denote the real and imaginary part of a complex number, respectively. $a(i,j) $ is the entry in the $i$-row and $j$-th column of matrix $\bf{A}$.

	\section{System Model and Problem Formulation}\label{SMPF}
	In this section, the considered MU-MIMO system model is firstly introduced, follow by a brief review of SLP method, based on which the joint transmit precoding and receive combining problem is formulated.
	
	\subsection{System Model}\label{SysMod}
	Consider a downlink multi-stream MU-MIMO system where the BS equipped with ${N_T}$ transmit antennas simultaneously transmits $L$ data streams to each user. Every user is equipped with ${N_R}$ antennas, which satisfies that $L \le N_R$, and the total number of transmit antennas satisfies that ${N_T} \ge KL$, where $K$ is the total number of users served by the BS. We focus on the downlink transmit precoding and receive combining design, where perfect CSI is assumed throughout the paper\cite{21,19,20,22}. The transmit symbol vector is assumed to be formed from a normalized PSK modulation constellation\cite{19}, where ${{\bf{s}}_k} \in \mathbb{C} {^{L \times 1}}$ is the transmit signal for the $k$-th user. The transmit precoding matrix for the $k$-th user can be expressed as ${{\bf{P}}_k} \in \mathbb{C} {^{N_T \times L}}$. Therefore, the transmit signal at the BS can be expressed as
	\begin{equation}\label{transmit signal}
		{\bf{x}} = {\bf{Ps}} = \sum\limits_{k = 1}^K {{{\bf{P}}_k}{{\bf{s}}_k}}, 
	\end{equation}
	where the ${\bf{P}} =\left[ {{{\bf{P}}_1},{{\bf{P}}_2}, \cdots ,{{\bf{P}}_K}} \right]\in \mathbb{C} {^{N_T \times KL}}$ and ${\bf{s}} = {\left[ {{\bf{s}}_1^T,{\bf{s}}_2^T, \cdots ,{\bf{s}}_K^T} \right]^T}\in \mathbb{C} {^{KL\times 1}}$ represent the transmit precoding matrix and symbol vector, respectively. Then, the received signal at the $k$-th user is given by
	\begin{equation}\label{receive signal}
		\begin{array}{l}
			{{\bf{y}}_k} = {{\bf{H}}_k}{\bf{x}} + {{\bf{n}}_k}\\
			{\kern 11pt}  = {{\bf{H}}_k}{{\bf{P}}_k}{{\bf{s}}_k} + {{\bf{H}}_k}\sum\limits_{i = 1,i \ne k}^K {{{\bf{P}}_i}{{\bf{s}}_i}}  + {{\bf{n}}_k}\\
			{\kern 11pt} = {{\bf{G}}_k}{{\bf{s}}_k} + \sum\limits_{i = 1,i \ne k}^K {{{\bf{G}}_{k,i}}{{\bf{s}}_i}}  + {{\bf{n}}_k},
		\end{array}
	\end{equation}
	where ${\bf{H}}_k \in \mathbb{C}{^{{N_R} \times {N_T}}}$ denotes the channel matrix between the $k$-th user and the BS, ${{\bf{n}}_k} \in \mathbb{C}{^{N_R \times 1}}  $ represents the zero mean circularly symmetric complex Gaussian noise vector with ${{\bf{n}}_k} \sim \mathcal{CN}\left( {{\bf{0}},{\sigma ^2}{\bf{I}}} \right)$. To simplify notation, we define ${{\bf{G}}_k} = {{\bf{H}}_k}{{\bf{P}}_k}$ and ${{\bf{G}}_{k,i} = {{\bf{H}}_k}{{\bf{P}}_i},k \ne i}$ as the effective user channel and interference channels, respectively.
	
	To correctly detect the received information symbols, the $k$-th user decodes the received signal with a receive combining matrix ${{\bf{W}}_k} \in \mathbb{C} {^{L \times N_R}}$, then the decoded signal can be expressed as
	\begin{equation}\label{decoded signal}
		\begin{array}{l}
			{{{\bf{\hat s}}}_k} = {{\bf{W}}_k}\left( {{{\bf{H}}_k}\sum\limits_{i = 1}^K {{{\bf{P}}_i}{{\bf{s}}_i}}  + {{\bf{n}}_k}} \right)\\
			{\kern 1pt} {\kern 1pt}  {\kern 1pt} {\kern 1pt} {\kern 1pt} {\kern 1pt} {\kern 1pt} {\kern 1pt} {\kern 1pt} {\kern 1pt}  = {{\bf{W}}_k}{{\bf{H}}_k}{{\bf{P}}_k}{{\bf{s}}_k} + {{\bf{W}}_k}{{\bf{H}}_k}\sum\limits_{i = 1,i \ne k}^K {{{\bf{P}}_i}{{\bf{s}}_i}}  + {{\bf{W}}_k}{{\bf{n}}_k},
		\end{array}
	\end{equation}
	
	\subsection{Constructive Interference}\label{CI}
\begin{figure}[!t]
		\centering
		\includegraphics[scale=0.5]{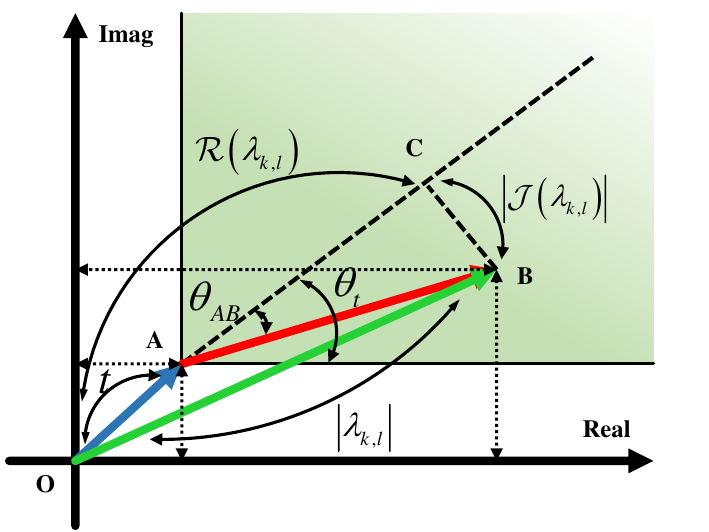}
		\caption{CI metric for QPSK modulation based on non-strict phase rotation. }
		\label{moxing}
	\end{figure}

	In order to enhance the comprehension of the subsequent problem formulation, a concise review of CI is introduced. As an illustrative example, Fig. \ref{moxing} shows the first quadrant of a QPSK constellation. The green shaded area is the CI region, which is defined as the interference responsible for displacing received signals beyond the detection thresholds\cite{21,19,20}. Treating the instantaneous interference as a source of the decoded signal, the communication performance can be enhanced by maximizing the distance between the received signal and decision boundaries. To provide an intuitive demonstration of this concept, we depict the scenario of non-strict phase rotation in Fig. \ref{moxing}.
	
	Without loss of the generality, we denote $|\overrightarrow {OA} | = t$ as the distance between the CI region and decision boundaries, which is the objective to be maximized. $\overrightarrow {OB} $ is denoted as the $l$-th noiseless decoded signal stream for the $k$-th user, which can be expressed as
	\begin{equation}\label{CI1}
		\overrightarrow {OB}  = {{\bf{w}}_{k,l}}{{\bf{H}}_k}{\bf{Ps}} = {\lambda _{k,l}}{s_{k,l}},
	\end{equation}
	where $ {{\bf{w}}_{k,l}}$ is the $l$-th row of combining matrix ${{\bf{W}}_k}$ and ${\lambda _{k,l}}$ is the introduced complex-valued scaling factor. Then, the node `C' is selected in half-line $\overrightarrow {OA}$ to make sure that $\overrightarrow {OC} $ and $\overrightarrow {CB}$ are perpendicular. Based on the Fig. \ref{moxing}, we can obtain
	\begin{equation}\label{CI2}
		\overrightarrow {OC}  = \mathcal{R}\left( {{\lambda _{k,l}}{s_{k,l}}} \right),\overrightarrow {CB}  = j \cdot  \mathcal{J}\left( {{\lambda _{k,l}}{s_{k,l}}} \right),|\overrightarrow {CB} | = \lambda _{k,l}^{\cal J},
	\end{equation}
	
	where unit `$ j$' denotes a $90^{\circ}$ phase rotation geometrically and  $\lambda _{k,l}^{\cal J} = {\cal J}\left( {{\lambda _{k,l}}} \right)$. Therefore, the length of the line segment $\overrightarrow {AC}$ is 
	\begin{equation}\label{CI3}
		|\overrightarrow {AC} | = {\cal R}\left( {{\lambda _{k,l}}{s_{k,l}}} \right) - t = \lambda _{k,l}^{\cal R} - t,
	\end{equation}
	where we denote $\lambda _{k,l}^{\cal R} = {\cal R}\left( {{\lambda _{k,l}}} \right)$. It can be observed that to have the received signal $\overrightarrow {OB}$ located in the CI region is equivalent to the following condition:
	\begin{equation}\label{CI4}
		\tan {\theta _{AB}} \le \tan {\theta _t},
	\end{equation}
	where $\theta _t = \frac{\pi}{M}$ is the constellation angle for an \textit{M}-PSK constellation, which leads to a more specific expression
	\begin{equation}\label{CI5}
		\left| {\lambda _{k,l}^{\cal J}} \right| - \left( {\lambda _{k,l}^{\cal R} - t} \right)\tan {\theta _t} \le 0,{\kern 1pt} \forall k \in {\cal K},\forall l \in {\cal L}.
	\end{equation}
	More detailed description can be found in\cite{131,19,22}.
	\subsection{Problem Formulation}\label{ProFor}
	Our objective is to maximize the distance between the constructive region and the detection thresholds, while adhering to constraints related to total transmit power budgets and non-strict CI constraints. Based on the aforementioned considerations, we can formulate the optimization problem as follows: 
	\begin{equation}\label{P1}
		\begin{array}{l}
			{\rm{\text{P1}}}:\mathop {\max }\limits_{{\bf{P}},{{\bf{W}}_k},t} {\kern 1pt} {\kern 1pt} {\kern 1pt} {\kern 1pt} t \vspace{1ex} \\ 
			{\kern 28pt}  s.t.{\kern 10pt} {{\bf{W}}_k}{{\bf{H}}_k}{\bf{Ps}} = {{\bf{\Lambda }}_k}{{\bf{s}}_k},\forall k \in {\cal K} \vspace{1ex}  \\
			{\kern 50pt}  \left| {\lambda _{k,l}^{\cal J}} \right| - \left( {\lambda _{k,l}^{\cal R} - t} \right)\tan {\theta _t} \le 0, \forall k \in {\cal K},\forall l \in {\cal L} \vspace{1ex} \\
			{\kern 50pt}  \left\| {{\bf{Ps}}} \right\|_2^2 \le {P_T} \vspace{1ex} \\
			{\kern 50pt} \left\| {{{\bf{W}}_k}} \right\|_2^2 \leq 1, \forall k \in {\cal K},
		\end{array}
	\end{equation}
	where ${{\bf{\Lambda }}_k} = diag\left( {{{\bm{\lambda }}_k}} \right)$ is a diagonal matrix and ${{\bm{\lambda }}_k} = {\left[ {{\lambda _{k,1}},{\lambda _{k,2}}, \cdots ,{\lambda _{k,L}}} \right]^T}$. In the formulated problem, the communication performance is maximized under the total transmit power budget ${P_T}$ and the non-strict CI constraints. We also enforce the power budget of the receive combining such that the objective is bounded and the problem is feasible, while we note that it will not affect the performance since ${{{\bf{W}}_k}}$ is also multiplied to the noise. 
	
	The proposed problem in (\ref{P1}) is distinct from the SEP constraint power minimization problem described in \cite{32} and the single-stream SLP problem for MU-MISO system discussed in \cite{19}. The objective here is to maximize the distance between the received signal and decision boundaries for each user's stream to ensure the best possible communication performance subject to the power budgets and SLP constraints. The additional variables introduced by the receive combining matrices ${\bf{W}}_k$ are coupled with the transmit precoding matrix $\bf{P}$. This coupling between the variables complicates the optimization process and makes it more computationally intensive, which makes the design of a computationally efficient algorithm for the optimization problem described in Equation (\ref{P1}) challenging. 
	
	\section{Joint Symbol-Level Transmit Precoding and Receive combining Design} \label{JSLTARBD}
	In this section, we focus on solving the formulated joint symbol level transmit precoding and receive combining design problem in (\ref{P1}), where the AO framework is adopted. The main reason for choosing this algorithm is twofold: First, the non-convex \text{P1} is split into two sub-problems with simpler structure by fixing each variable as a constant so that $\bf P$, ${\bf{W}}_k$ can be optimized separately in each sub-problem. Second, the precoding matrix $\bf P$ and receive combining matrix ${\bf{W}}_k$ are loosely coupled in the constraints, a monotonic convergence of the objective value can be achieved so that a stationary point solution of \text{P1} can be obtained by using the AO algorithm. When the variable ${\bf{W}}_k$ is fixed for optimizing $\bf P$, the formulated sub-problem is convex and the closed-form solution can be obtained. When updating the ${\bf{W}}_k$ with the optimized $\bf P$, the closed-form solution also can be obtained.
	
	\subsection{Optimization on Transmit Precoding Matrix ${\bf{P}}$}\label{P}
	We first present the optimization of the transmit precoding matrix {\bf{P}} for the given receive combining matrix ${{\bf{W}}_k}$. In particular, the sub-problem is similar to the formulated problem in work \cite{19} and the closed-form solution can be obtained by analyzing the Lagrangian and KKT conditions. When the ${{\bf{W}}_k}$ is fixed, \text{P1} can be transformed into
	\begin{equation}\label{P2}
		\begin{array}{l}
			{\rm{\text{P2}}}:\mathop {\max }\limits_{{\bf{P}},t} {\kern 1pt} {\kern 1pt} {\kern 1pt} {\kern 1pt} t \vspace{1ex} \\ 
			{\kern 20pt}  s.t.{\kern 10pt} {{\bf{F}}_k}{\bf{Ps}} = {{\bf{\Lambda }}_k}{{\bf{s}}_k},\forall k \in {\cal K} \vspace{1ex}  \\
			{\kern 42pt}  \left| {\lambda _{k,l}^{\cal J}} \right| - \left( {\lambda _{k,l}^{\cal R} - t} \right)\tan {\theta _t} \le 0, \forall k \in {\cal K},\forall l \in {\cal L} \vspace{1ex} \\
			{\kern 42pt}  \left\| {{\bf{Ps}}} \right\|_2^2 \le {P_T}, \\
		\end{array}
	\end{equation}
	where ${{\bf{F}}_k} = {{\bf{W}}_k}{{\bf{H}}_k} \in \mathbb{C} {^{L \times Nt}}$. According to the work in \cite{19}, the closed-form solution of ${\bf{P}}$ can be obtained as 
	\begin{equation}\label{closed-form W}
		{\bf{P}} = \frac{1}{{KL}}{{\bf{F}}^H}{\left( {{\bf{F}}{{\bf{F}}^H}} \right)^{ - 1}}diag\left\{ {\sqrt {\frac{{{P_T}}}{{{{{\bf{\hat u}}}^T}{\bf{V\hat u}}}}} {\bf{U}}{{{\bf{\hat T}}}^{ - 1}}{{\bf{S}}^T}{\bf{\hat u}}} \right\}{\bf{s\hat s}},
	\end{equation}
	where ${\bf{\hat s}} = \left[ {\dfrac{1}{{{s_1}}},\dfrac{1}{{{s_2}}}, \cdots ,\dfrac{1}{{{s_{KL}}}}} \right]$, ${\bf{F}} = {\left[ {{\bf{F}}_1^T,{\bf{F}}_2^T, \cdots ,{\bf{F}}_K^T} \right]^T}$ and ${\bf{U}} = \left[ {\begin{array}{*{20}{c}}
			{\bf{I}}&{j \cdot {\bf{I}}}
	\end{array}} \right]$. ${\bf{\hat u}}$ is the dual variables and expressed as
	\begin{equation}\label{variable1}
		\begin{array}{l}
			{\bf{\hat u}} = {\left[ {{\hat \mu _{1,1}}, \cdots ,{\hat \mu _{k,l}}, \cdots ,{\hat \mu _{K,L}},{\hat \nu _{1,1}}, \cdots ,{\hat \nu _{k,l}}, \cdots {\hat \nu _{K,L}}} \right]^T},\\
		\end{array}
	\end{equation}
	which can be obtained by solving a simpler QP optimization problem 
	\begin{equation}\label{P3}
		\begin{array}{l}
			{\text{P3}}:\mathop {\min }\limits_{\bf{\hat u}} {{\bf{\hat u}}^T}{\bf{V \hat u}} \vspace{1ex}\\
			{\kern 20pt}s.t.{\kern 5pt} {{\bf{1}}^T}{\bf{ \hat u}} = 1 \vspace{1ex}\\
			{\kern 40pt} {u_{i}} \ge 0,\forall i \in \left\{ {1,2, \cdots ,2KL} \right\},
		\end{array}
	\end{equation}
	where ${\bf{V}} = {\bf{S}}{{\bf{\hat T}}^{ - 1}}{{\bf{S}}^T}$, ${\bf{T}} = diag({{\bf{s}}^H}){\left( {{\bf{F}}{{\bf{F}}^H}} \right)^{ - 1}}diag({\bf{s}})$. Furthermore, ${\bf{\hat T}}$ and $\bf{S}$ are expressed as
	\begin{equation}\label{decompose}
		{\bf{\hat T}} = \left[ {\begin{array}{*{20}{c}}
				{{\cal R}\left( {\bf{T}} \right)}&{ - {\cal J}\left( {\bf{T}} \right)}\\
				{{\cal J}\left( {\bf{T}} \right)}&{{\cal R}\left( {\bf{T}} \right)}
		\end{array}} \right],
	\end{equation}
	\begin{equation}\label{variable2}
		{\bf{S}} = \left[ {\begin{array}{*{20}{c}}
				{\bf{I}}&{ - \dfrac{1}{{\tan {\theta _t}}} \cdot {\bf{I}}}\\
				{\bf{I}}&{\dfrac{1}{{\tan {\theta _t}}} \cdot {\bf{I}}}
		\end{array}} \right].
	\end{equation}
	
	Moreover, \text{P3} is a typical QP optimization problem compared with the original problem in (\ref{P2}), and it has already been shown in the existing literature that this optimization can be more efficiently solved\cite{19}. It can be observed that the numbers of optimized variables of \text{P2} and \text{P3} are ${N_T}KL$ and $2KL$, respectively, and obviously, the variable size is reduced.
	\subsection{Optimization on Receive Combining Matrix ${{\bf{W}}_k}$}\label{Wk}
	Next, we present the optimization of receive combining matrix ${{\bf{W}}_k}$ for the given transmit precoding matrix $\bf P$, which can be formulated as
	\begin{equation}\label{P4}
		\begin{array}{l}
			{\text{P4}}:\mathop {\min }\limits_{{{\bf{W}}_k},t}- t \vspace{1ex}\\
			{\kern 23pt} s.t.{\kern 3pt} {{\bf{W}}_k}{{\bf{r}}_k} = {{\bf{\Lambda }}_k}{{\bf{s}}_k},\forall k \in {\cal K} \vspace{1ex}\\
			{\kern 38pt}  \left| {\lambda _{k,l}^{\cal J}} \right| - \left( {\lambda _{k,l}^{\cal R} - t} \right)\tan {\theta _t} \le 0, \forall k \in {\cal K},\forall l \in {\cal L} \vspace{1ex}\\
			{\kern 38pt}  \left\| {{{\bf{W}}_k}} \right\|_2^2 \leq 1, \forall k \in {\cal K},
		\end{array}
	\end{equation}
	where ${{\bf{r}}_k} = {{\bf{H}}_k}{\bf{Ps}}$ is the noiseless received signal for $k$-th user. The Lagrangian function of \text{P4} is
	\begin{equation}\label{L3}
		\begin{array}{l}
			{\cal L}\left( {{{\bf{w}}_{k,l}},t,{\xi _k},{\delta _{k,l}},{\gamma _{k,l}}} \right) \vspace{1ex}\\
			=  - t + \sum\limits_{k = 1}^K {{\xi _k}\left( {\sum\limits_{l = 1}^L {{{\bf{w}}_{k,l}}{\bf{w}}_{k,l}^H}  - 1} \right)} \vspace{1ex}\\
			+ \sum\limits_{k = 1}^K {\sum\limits_{l = 1}^L {{\delta _{k,l}}\left( {{{\bf{w}}_{k,l}}{{\bf{r}}_k} - {\lambda _{k,l}}{s_{k,l}}} \right)} } \vspace{1ex} \\
			+ \sum\limits_{k = 1}^K {\sum\limits_{l = 1}^L {{\gamma _{k,l}}\left[ {\left| {\lambda _{k,l}^{\cal J}} \right| - \left( {\lambda _{k,l}^{\cal R} - t} \right)\tan {\theta _t}} \right]} } ,
		\end{array}
	\end{equation}
	where ${{\bf{w}}_{k,l}}$ is the $l$-th row of ${{\bf{W}}_k}$, and ${\xi _k} \geq 0$, ${\delta _{k,l}}$ and ${\gamma _{k,l}} \geq 0$ are the dual variables. Each ${\delta _{k,l}}$ may be complex as it is the dual variable with respect to the equality constraint. Based on the Lagrangian in (\ref{L3}), the KKT conditions for optimality can be obtained as
	\begin{subequations}\label{KKT2}
		\begin{align}
			&\label{c1}\frac{{\partial {\cal L}}}{{\partial {{\bf{w}}_{k,l}}}} = {\delta _{k,l}}{{\bf{r}}_k} + {\xi _k}{\bf{w}}_{k,l}^H = {\bf{0}},\forall k \in {\cal K},\forall l \in {\cal L}\\  
			&\label{c2}\frac{{\partial {\cal L}}}{{\partial t}} = \sum\limits_{k = 1}^K {\sum\limits_{l = 1}^L {{\gamma _{k,l}}} }  - 1 = 0\\  
			&\label{c3}{{{\xi _k}\left( {\sum\limits_{l = 1}^L {{{\bf{w}}_{k,l}}{\bf{w}}_{k,l}^H}  - 1} \right)}}=0,\forall k \in {\cal K},\forall l \in {\cal L}\\
			&\label{c4}{{{\bf{w}}_{k,l}}{{\bf{r}}_k} - {\lambda _{k,l}}{s_{k,l}}}=0,\forall k \in {\cal K},\forall l \in {\cal L}\\
			&\label{c5}{{\gamma _{k,l}}\left( {\left| {\lambda _{k,l}^{\cal J}} \right| - \left( {\lambda _{k,l}^{\cal R} - t} \right)\tan {\theta _t}} \right)}=0,\forall k \in {\cal K},\forall l \in {\cal L}.
		\end{align}
	\end{subequations}
	Based on (\ref{c1}), we can obtain that ${\xi _k \ne 0}$, and with the fact that ${\xi _k} \geq 0$ we can further have ${\xi _k > 0}$. Then, the ${{\bf{p}}_{k,l}}$ can be expressed as
	\begin{equation}\label{p_k,l1}
		\begin{array}{l}
			{\bf{w}}_{k,l}^H =  - \dfrac{{{\delta _{k,l}}{{\bf{r}}_k}}}{{{\xi _k}}} = {\zeta _{k,l}}{{\bf{r}}_k} \vspace{1ex}\\
			\Rightarrow {{\bf{w}}_{k,l}} = \zeta _{k,l}^*{\bf{r}}_k^H,\forall k \in {\cal K},\forall l \in {\cal L},
		\end{array}
	\end{equation}
	where ${\zeta _{k,l}}=-\dfrac{{{\delta _{k,l}}}}{{{\xi _k}}}$. Then, with each ${\bf{w}}_{k,l}$ is obtained, the receive combining matrix for $k$-th user can be further expressed in a matrix form as
	\begin{equation}\label{P_k}
		{{\bf{W}}_k} = \left[ {\begin{array}{*{20}{c}}
				{{{\bf{w}}_{k,1}}}\\
				{{{\bf{w}}_{k,2}}}\\
				\vdots \\
				{{{\bf{w}}_{k,L}}}
		\end{array}} \right] = \left[ {\begin{array}{*{20}{c}}
				{\zeta _{k,1}^*{\bf{r}}_k^H}\\
				{\zeta _{k,2}^*{\bf{r}}_k^H}\\
				\vdots \\
				{\zeta _{k,L}^*{\bf{r}}_k^H}
		\end{array}} \right] = \left[ {\begin{array}{*{20}{c}}
				{\zeta _{k,1}^*}\\
				{\zeta _{k,2}^*}\\
				\vdots \\
				{\zeta _{k,L}^*}
		\end{array}} \right]{\bf{r}}_k^H = {\bm{\zeta}{^*_k} \bf r}_k^H,
	\end{equation}
	where we have ${\bm{\zeta}}{_k}={\left[ {{\zeta _{k,1}},{\zeta _{k,2}}, \cdots ,{\zeta _{k,L}}} \right]^T}$. Then, the decoded signal can be further expressed as 
	\begin{equation}\label{zeta}
		\begin{array}{l}
			{{\bf{W}}_k}{{\bf{r}}_k} = {{\bf{\Lambda }}_k}{{\bf{s}}_k} \vspace{1ex}\\
			\Rightarrow {\bm{\zeta}}{^*_k} {\bf{r}}{_k^H}{{\bf{r}}_k} = {{\bf{\Lambda }}_k}{{\bf{s}}_k} \vspace{1ex}\\
			\Rightarrow {\bm{\zeta}}{^*_k} = \dfrac{1}{{{\bf{r}}_k^H{{\bf{r}}_k}}}{{\bf{\Lambda }}_k}{{\bf{s}}_k}.
		\end{array}
	\end{equation}
	With (\ref{zeta}), we can obtain the structure of the optimal receive combining matrix ${\bf W}{_k}$ as a function of $ {{\bf{\Lambda }}_k}$ as
	\begin{equation}\label{P_k1}
		{{\bf{W}}_k} = {\bm{\zeta}{^*_k} \bf r}_k^H = \dfrac{1}{{{\bf{r}}_k^H{{\bf{r}}_k}}}{{\bf{\Lambda }}_k}{{\bf{s}}_k}{\bf{r}}_k^H.
	\end{equation}
	With the fact that  ${\xi _k > 0}$, based on (\ref{c3}) we can obtain that the normalized requirement is strictly active, which leads to
	\begin{equation}\label{normalized}
		\begin{array}{l}
			\left\| {{{\bf{W}}_k}} \right\|_2^2 = 1\vspace{1ex}\\
			\Rightarrow {\rm{Tr}}\left( {{{\bf{W}}_k}{\bf{W}}_k^H} \right) = 1\vspace{1ex}\\
			\Rightarrow {\rm{Tr}}\left( {\dfrac{1}{{{{\left( {{\bf{r}}_k^H{{\bf{r}}_k}} \right)}^2}}}{{\bf{\Lambda }}_k}{{\bf{s}}_k}{\bf{r}}_k^H{{\bf{r}}_k}{\bf{s}}_k^H{\bf{\Lambda }}_k^H} \right) = 1\vspace{1ex}\\
			\Rightarrow \dfrac{1}{{\left( {{\bf{r}}_k^H{{\bf{r}}_k}} \right)}}{\bf{s}}_k^H{\bf{\Lambda }}_k^H{{\bf{\Lambda }}_k}{{\bf{s}}_k} = 1\vspace{1ex}\\
			\Rightarrow \dfrac{1}{{\left( {{\bf{r}}_k^H{{\bf{r}}_k}} \right)}}{\bm{\lambda }}_k^Hdiag\left( {{\bf{s}}_k^H} \right)diag\left( {{{\bf{s}}_k}} \right){{\bm{\lambda }}_k} = 1\vspace{1ex}\\
			\Rightarrow \dfrac{1}{{\left( {{\bf{r}}_k^H{{\bf{r}}_k}} \right)}}{\bm{\lambda }}_k^H{{\bm{\lambda }}_k} = 1,
		\end{array}
	\end{equation}
	then we define ${{\bm{\hat \lambda }}_k} = {\left[ {{\cal R}\left( {{\bm{\lambda }}_k^T} \right),{\cal J}\left( {{\bm{\lambda }}_k^T} \right)} \right]^T}$ and we could have
	\begin{equation}\label{lambda}
		\begin{array}{l}
			\dfrac{1}{{\left( {{\bf{r}}_k^H{{\bf{r}}_k}} \right)}}{\bm{\hat \lambda }}_k^T{{{\bm{\hat \lambda }}}_k} = 1\vspace{1ex}\\
			\Rightarrow {\bm{\hat \lambda }}_k^T{{{\bm{\hat \lambda }}}_k} = \left( {{\bf{r}}_k^H{{\bf{r}}_k}} \right) = {P_k},
		\end{array}
	\end{equation}
	where ${{{\bf{r}}_k}}$ is the known noiseless received signal, which makes ${P_k}$ constant in the problem. Therefore, \text{P4} can be transformed into 
	\begin{equation}\label{P5}
		\begin{array}{l}
			{\text{P5}}:\mathop {\min }\limits_{{\bm{\hat \lambda }_k},t} {\kern 1pt}  - t \vspace{1ex}\\
			{\kern 21pt}s.t.{\kern 5pt} {\bm{\hat \lambda }}_k^T{{{\bm{\hat \lambda }}}_k} - {P_k} = 0,\forall k \in {\cal K} \vspace{1ex}\\
			{\kern 38pt}  \dfrac{{\lambda _{k,l}^{\cal J}}}{{\tan {\theta _t}}} + t - \lambda _{k,l}^{\cal R} \le 0,{\kern 1pt} \forall k \in {\cal K},\forall l \in {\cal L} \vspace{1ex}\\
			{\kern 38pt}  - \dfrac{{\lambda _{k,l}^{\cal J}}}{{\tan {\theta _t}}} + t - \lambda _{k,l}^{\cal R} \le 0,\forall k \in {\cal K},\forall l \in {\cal L},
		\end{array}
	\end{equation}
	where the CI constraint is transformed into two separate constraints. We analyze \text{P5} with Lagrangian and KKT conditions, where the Lagrangian function is constructed as
	\begin{equation}\label{L4}
		\begin{array}{l}
			{\cal L}\left( {{{{\bm{\hat \lambda }}}_k},t,{{\tilde \alpha }_k},{{\tilde \mu }_{k,l}},{{\tilde \nu }_{k,l}}} \right) \vspace{1ex}\\
			=  - t + \sum\limits_{k = 1}^K {{{\tilde \alpha }_k}\left( {{\bm{\hat \lambda }}_k^T{{{\bm{\hat \lambda }}}_k} - {P_k}} \right)} \vspace{1ex}\\
			+ \sum\limits_{k = 1}^K {\sum\limits_{l = 1}^L {{{\tilde \mu }_{k,l}}\left( {\dfrac{{\lambda _{k,l}^{\cal J}}}{{\tan {\theta _t}}} + t - \lambda _{k,l}^{\cal R}} \right)} } \vspace{1ex}\\
			+ \sum\limits_{k = 1}^K {\sum\limits_{l = 1}^L {{{\tilde \nu }_{k,l}}\left( { - \dfrac{{\lambda _{k,l}^{\cal J}}}{{\tan {\theta _t}}} + t - \lambda _{k,l}^{\cal R}} \right)} } \vspace{1ex}\\
			= \left[ {\sum\limits_{k = 1}^K {\sum\limits_{l = 1}^L {\left( {{{\tilde \mu }_{k,l}} + {{\tilde \nu }_{k,l}}} \right)}  - 1} } \right]t + \sum\limits_{k = 1}^K {{{\tilde \alpha }_k}\left( {{\bm{\hat \lambda }}_k^T{{{\bm{\hat \lambda }}}_k} - {P_k}} \right)} \vspace{1ex}\\
			- \sum\limits_{k = 1}^K {\sum\limits_{l = 1}^L {\left( {{{\tilde \mu }_{k,l}} + {{\tilde \nu }_{k,l}}} \right)\lambda _{k,l}^{\cal R}} }  + \sum\limits_{k = 1}^K {\sum\limits_{l = 1}^L {\left( {{{\tilde \mu }_{k,l}} - {{\tilde \nu }_{k,l}}} \right)\dfrac{{\lambda _{k,l}^{\cal J}}}{{\tan {\theta _t}}}} } ,
		\end{array}
	\end{equation}
	where ${{\tilde \alpha }_k}$, ${{\tilde \mu }_{k,l}} \geq 0 $ and ${{\tilde \nu }_{k,l}} \geq 0, \forall k \in {\cal K},\forall l \in {\cal L}$ are the dual variables and ${{\tilde \alpha }_k}$ may be complex as it is the dual variable with respect to the equality constraint. In order to transform the Lagrangian function into a more compact form, we define that ${{\bf{\tilde u}}_k} = {\left[ {{{\tilde \mu }_{k,1}}, \cdots ,{{\tilde \mu }_{k,L}},{{\tilde \nu }_{k,1}}, \cdots {{\tilde \nu }_{k,L}}} \right]^T}$ and $\bf{S}$ is in (\ref{variable2}), then the Lagrangian function can be expressed as
	\begin{equation}\label{L5}
		\begin{array}{l}
			{\cal L}\left( {{{{\bm{\hat \lambda }}}_k},t,{{\tilde \alpha }_k},{{{\bf{\tilde u}}}_k}} \right)\vspace{1ex} \\
			= \sum\limits_{k = 1}^K {{{\tilde \alpha }_k}\left( {{\bm{\hat \lambda }}_k^T{{{\bm{\hat \lambda }}}_k} - {P_k}} \right)}  + \left( {\sum\limits_{k = 1}^K {{{\bf{1}}^T}{{{\bf{\tilde u}}}_k}}  - 1} \right)t - \sum\limits_{k = 1}^K {{\bf{\tilde u}}_k^T{\bf{S}}{{{\bm{\hat \lambda }}}_k}}.
		\end{array}
	\end{equation}
	Based on the Lagrangian in (\ref{L5}), the KKT conditions for optimality can be obtained as
	\begin{subequations}\label{KKT3}
		\begin{align}
			&\label{d1}\dfrac{{\partial {\cal L}}}{{\partial t}} = \sum\limits_{k = 1}^K {{{\bf{1}}^T}{{{\bf{\tilde u}}}_k}}  - 1 = 0\\  
			&\label{d2}\dfrac{{\partial {\cal L}}}{{\partial {{{\bm{\hat \lambda }}}_k}}} = 2{{\tilde \alpha }_k}{{{\bm{\hat \lambda }}}_k} - {{\bf{S}}^T}{{{\bf{\tilde u}}}_k} = {\bf{0}},\forall k \in {\cal K} \\  
			&\label{d3}{{\bm{\hat \lambda }}_k^T{{{\bm{\hat \lambda }}}_k} - {P_k}} = 0 ,\forall k \in {\cal K} \\
			&\label{d4}{{{\tilde \mu }_{k,l}}\left( {\dfrac{{\lambda _{k,l}^{\cal J}}}{{\tan {\theta _t}}} + t - \lambda _{k,l}^{\cal R}} \right)}=0,\forall k \in {\cal K},\forall l \in {\cal L}\\
			&\label{d5}{{{\tilde \nu }_{k,l}}\left( {- \dfrac{{\lambda _{k,l}^{\cal J}}}{{\tan {\theta _t}}} + t - \lambda _{k,l}^{\cal R}} \right)}=0,\forall k \in {\cal K},\forall l \in {\cal L},
		\end{align}
	\end{subequations}
	where ${{\tilde \alpha }_k} \ne 0$ based on (\ref{d2}), then we have
	\begin{equation}\label{lambdak}
		{{{\bm{\hat \lambda }}}_k} = \frac{1}{{2{{\tilde \alpha }_k}}}{{\bf{S}}^T}{{{\bf{\tilde u}}}_k}.
	\end{equation}
	By substituting the expression of ${{{\bm{\hat \lambda }}}_k}$ in (\ref{lambdak}) into (\ref{d3}), we can further express ${{\tilde \alpha }_k}$ as a function of the dual vector $ {{{\bf{\tilde u}}}_k}$, given by
	\begin{equation}\label{uk}
		\begin{array}{l}
			{\bm{\hat \lambda }}_k^T{{{\bm{\hat \lambda }}}_k} = {P_k} \vspace{1ex}\\
			\Rightarrow {\left( \dfrac{1}{{2{{\tilde \alpha }_k}}}{{\bf{S}}^T}{{{\bf{\tilde u}}}_k} \right)^T}\left( \dfrac{1}{{2{{\tilde \alpha }_k}}}{{\bf{S}}^T}{{{\bf{\tilde u}}}_k} \right) = {P_k} \vspace{1ex}\\
			\Rightarrow {{\tilde \alpha }_k} = \sqrt {\dfrac{{{\bf{\tilde u}}_k^T{\bf{S}}{{\bf{S}}^T}{{{\bf{\tilde u}}}_k}}}{{4{P_k}}}}.
		\end{array}
	\end{equation}
	
	The convex optimization \text{P5} in (\ref{P5}) satisfies the Slater's condition and the dual gap is zero\cite{33}. Therefore, With the obtained ${{{\bm{\hat \lambda }}}_k}$ in (\ref{lambdak}) and ${{\tilde \alpha }_k}$ in (\ref{uk}), \text{P5} can be solved by solving its dual problem, which is given by
	\begin{equation}\label{L6}
		\begin{array}{l}
			{\cal U} = \mathop {\max }\limits_{{{{\bf{\tilde u}}}_k},{{\tilde \alpha }_k}} \mathop {\min }\limits_{{{{\bm{\hat \lambda }}}_k},t} {\cal L}\left( {{{{\bm{\hat \lambda }}}_k},t,{{\tilde \alpha }_{k}},{{{\bf{\tilde u}}}_k}} \right)\vspace{1ex}\\
			{\kern 8pt}= \mathop {\max }\limits_{{{{\bf{\tilde u}}}_k},{{\tilde \alpha }_k}} \sum\limits_{k = 1}^K {{{\tilde \alpha }_k}\left( {{\bm{\hat \lambda }}_k^T{{{\bm{\hat \lambda }}}_k} - {P_k}} \right)}  - \sum\limits_{k = 1}^K {{\bf{\tilde u}}_k^T{\bf{S}}{{{\bm{\hat \lambda }}}_k}} \vspace{1ex}\\
			{\kern 8pt}= \mathop {\max }\limits_{{{{\bf{\tilde u}}}_k},{{\tilde \alpha }_k}} \sum\limits_{k = 1}^K {{ - {{\tilde \alpha }_k}{P_k} - {\bf{\tilde u}}_k^T{\bf{S}}{{{\bm{\hat \lambda }}}_k}}} \vspace{1ex}\\
			{\kern 18pt}+{{\tilde \alpha }_k} {\left( \dfrac{1}{{2{{\tilde \alpha }_k}}}{{\bf{S}}^T}{{{\bf{\tilde u}}}_k} \right)^T}\left( \dfrac{1}{{2{{\tilde \alpha }_k}}}{{\bf{S}}^T}{{{\bf{\tilde u}}}_k} \right) \vspace{1ex}\\
			{\kern 8pt}= \mathop {\max }\limits_{{{{\bf{\tilde u}}}_k}} {\kern 1pt} {\kern 1pt}  - \sum\limits_{k = 1}^K { \sqrt {{P_k}{\bf{\tilde u}}_k^T{\bf{S}}{{\bf{S}}^T}{{\bf{\tilde u}}_k}}} ,
		\end{array}
	\end{equation}
	then the dual problem $\hat {\cal U}$ is equivalent to the following optimization problem
	\begin{equation}\label{P6}
		\begin{array}{l}
			{\text{P6}}: \mathop {\min }\limits_{{\bf{\tilde u}}} \sum\limits_{k = 1}^K {\sqrt {{P_k}} {{\left\| {{\bf{\tilde u}}_k^T{\bf{S}}} \right\|}_2}} \vspace{1ex}\\
			{\kern 20pt}  s.t.{\kern 1pt} \sum\limits_{k = 1}^K {{{\bf{1}}^T}{{{\bf{\tilde u}}}_k} = 1} \vspace{1ex}\\
			{\kern 35pt}  {{\tilde u}_{k,l}} \ge 0,\forall k \in \left\{ {1,2, \cdots K} \right\},\forall l \in \left\{ {1,2, \cdots 2L} \right\}
		\end{array}
	\end{equation}
	where the first constraint comes from (\ref{d1}).
	
	Based on the above analysis and transformations, the original receive combining matrix optimization problem in (\ref{P5}) can be solved by solving \text{P6} in (\ref{P6}). More specifically, the closed-form receive combining matrix can be obtained after obtaining the ${{{\bm{\hat \lambda }}}_k}$ in (\ref{lambdak}) and ${{\tilde \alpha }_k}$ in (\ref{uk}), whose expression is given by
	\begin{equation}\label{closedformP}
		{{\bf{W}}_k} = \frac{1}{{{\bf{r}}_k^H{{\bf{r}}_k}}}diag\left( {\sqrt {\frac{{{P_k}}}{{{\bf{\tilde u}}_k^T{\bf{S}}{{\bf{S}}^T}{{{\bf{\tilde u}}}_k}}}} } {{\bf{S}}^T}{{{\bf{\tilde u}}}_k} \right){{\bf{s}}_k}{\bf{r}}_k^H.
	\end{equation}
	
	With the above derivation from the (\ref{P5}) to (\ref{P6}), the original sub-problem is transformed into a easier convex problem and can be efficiently solved by using CVX tool\cite{33}. Obviously, it can be observed that the numbers of variables of the (\ref{P5}) and (\ref{P6}) are ${N_R}KL$ and $2KL$ ($N_R \geq 2$ in our considered MIMO system), which indicates that the proposed solution can reduce more complexity with the increase in the number of receive antennas.
	
	\subsection{Alternating Optimization Algorithm}\label{IRC1}
	In Section \ref{P} and Section \ref{Wk}, the original problem (\ref{P1}) has been transformed into two simpler convex sub-problems, i.e., (\ref{P4}) and (\ref{P6}). Based on the above, we present the AO algorithm to solve \text{P1}, which is summarized in \textbf{Algorithm \ref{al1}}.
	
	\begin{algorithm}[!h]
		\caption{Proposed AO algorithm for solving problem (\ref{P1})}
		\label{al1}
		\begin{algorithmic}
			\State ${\bf Input:}$ ${\bf{H}}$, ${\bf{s}}$, and $\kappa$.\vspace{0.5ex}
			\State ${\bf Output:}$ ${\bf{P}}$, ${\bf{W}}_k$.\vspace{0.5ex}
			\State ${\bf Initialization:}$  Set the variable ${\bf{P}}^{(1)}$ with BD algorithm, $t^{(1)}=0$, $ \kappa=10{^{-5}}$, and iteration index $ \textit{n}=1 $.\vspace{0.5ex}
			\While {$\delta > \kappa $} \vspace{0.5ex}
			\State Obtain ${\bf{\tilde u}}$ via (\ref{P6}) \vspace{0.5ex}
			\State Obtain ${{\bf{W}}_{k}^{(n+1)}} $  via (\ref{closedformP})\vspace{0.5ex}
			\State Obtain ${{\bf{\hat u}}}$ via (\ref{P3}) \vspace{0.5ex}
			\State Obtain ${{\bf{P}}^{(n+1)}}$ and $t^{(n+1)}$ via (\ref{closed-form W}) and (\ref{P2}) \vspace{0.5ex}
			\State $\delta  = \left| t^{(n+1)}-t^{(n)} \right|$
			\State $n=n+1$
			\EndWhile
			\State ${\bf P}={\bf P}^{\left( {n} \right)}$, ${{\bf{W}}_k}={\bf{W}}_k^{\left( {n} \right)}$.
		\end{algorithmic}
	\end{algorithm}
	First, we initialize the precoding matrix ${\bf{P}}^{(1)}$ with the BD algorithm. For given ${\bf{P}}^{(n)}$, the solution of ${\bf{\tilde u}}$ can be obtained by solving the convex problem (\ref{P6}) and the solution of the combining matrix ${{\bf{W}}_{k}^{(n+1)}} $ can be calculated directly in the (\textit{n}+1)-th iteration by using (\ref{closedformP}). With the updated ${{\bf{W}}_{k}^{(n+1)}} $, ${{\bf{\hat u}}}$ can be obtained by solving the QP problem (\ref{P3}), the solution of precoding matrix ${\bf{P}}^{(n+1)}$ and optimized objective $t$ are obtained by using (\ref{closed-form W}) and (\ref{P2}). The convergence and computational complexity analysis of the proposed algorithm is in Section \ref{CCCA}.
	
	\section{Symbol-Level Precoding with RIRC Receiver} \label{SLPWIR}
	One major challenge of the proposed joint transmit precoding and receive combining optimization in Section \ref{JSLTARBD} is that the design of receive combining matrix is related to the transmit signal, which is highly impractical. To address this challenge, the IRC receiver is firstly proposed in this section. However, the rank-one SLP matrix poses a difficulty in direct utilization. Subsequently, we introduce the RIRC receiver, which allows for iterative optimization of the transmit precoding matrix and receive combining matrix.
	\subsection{Interference Rejection Combiner}\label{IRC2}
	Traditionally, in order to correctly detect the received signal, the interference of inter-user in MU-MIMO system must be suppressed and hence the interference aware receivers are used widely. Especially, the IRC receiver helps to improve the interference rejection capabilities at the receiver end by exploiting the structure of the interference\cite{35}. Next, a brief introduction of the IRC receiver is given.
	
	Based on the signal model in (\ref{receive signal}), the IRC matrix for $k$-th user of MU-MIMO system can be expressed as
	\begin{equation}\label{IRC}
		{{\bf{W}}_k} = \frac{{{\bf{G}}_k^H{\bf{R}}_\eta ^{ - 1}}}{{{\bf{G}}_k^H{\bf{R}}_\eta ^{ - 1}{{\bf{G}}_k}}} = {\left( {{\bf{G}}_k^H{\bf{R}}_\eta ^{ - 1}{{\bf{G}}_k}} \right)^{ - 1}}{\bf{G}}_k^H{\bf{R}}_\eta ^{ - 1},
	\end{equation}
	where ${\bm{\eta }} = \sum\nolimits_{i = 1,i \ne k}^K {{{\bf{G}}_{k,i}}{{\bf{s}}_i}}  + {{\bf{n}}_k}$ is the sum of interference and noise vector for $k$-th user, and the covariance matrix of ${\bm{\eta }} $ can be expressed as
	\begin{equation}\label{RETA}
		\begin{array}{l}
			{{\bf{R}}_{\bm{\eta }}} =\mathbb{E} \left\{ {\left[ {{\bm{\eta }} -\mathbb{E} \left( {\bm{\eta }} \right)} \right]{{\left[ {{\bm{\eta }} - \mathbb{E} \left( {\bm{\eta }} \right)} \right]}^H}} \right\} =\mathbb{E} \left\{ {{\bm{\eta }}{{\bm{\eta }}^H}} \right\}\\
			{\kern 15pt}= \sum\limits_{i = 1,i \ne k}^K {{{\bf{G}}_{k,i}}\mathbb{E} \left\{ {{{\bf{s}}_i}{\bf{s}}_i^H} \right\}{\bf{G}}_{k,i}^H}  + \mathbb{E} \left\{ {{{\bf{n}}_i}{\bf{n}}_i^H} \right\}\\
			{\kern 15pt}= \sum\limits_{i = 1,i \ne k}^K {{{\bf{G}}_{k,i}}{\bf{G}}_{k,i}^H}  + {\sigma ^2}{\bf{I}}.
		\end{array}
	\end{equation}
	\subsection{SLP Matrix Analysis}\label{PF}
	When the IRC receiver is used to detect the received signal, the AO algorithm can also be used to solve the problem. First, the precoding matrix is initialized by using BD method\cite{25}. Then, the IRC matrix  can be ${{\bf{W}}_k}$ obtained by using (\ref{IRC}). Second, with the given IRC matrix, the precoding matrix can be optimized by solving the following problem
	\begin{equation}\label{P7}
		\begin{array}{l}
			{\rm{\text{P7}}}:\mathop {\max }\limits_{{\bf{P}},t} {\kern 1pt} {\kern 1pt} {\kern 1pt} {\kern 1pt} t \vspace{1ex} \\ 
			{\kern 20pt}  s.t.{\kern 10pt} {{\bf{F}}_k}{\bf{Ps}} = {{\bf{\Lambda }}_k}{{\bf{s}}_k},\forall k \in {\cal K} \vspace{1ex}  \\
			{\kern 42pt}  \left| {\lambda _{k,l}^{\cal J}} \right| - \left( {\lambda _{k,l}^{\cal R} - t} \right)\tan {\theta _t} \le 0, \forall k \in {\cal K},\forall l \in {\cal L} \vspace{1ex} \\
			{\kern 42pt}  \left\| {{\bf{Ps}}} \right\|_2^2 \le {P_T}, \\
		\end{array}
	\end{equation}
	then we can employ the same methodology discussed in Section \ref{P} to obtain the precoding matrix ${\bf{P}}$ by using (\ref{closed-form W}). However, the SLP matrix cannot be directly used to update the receive combining matrix due to the fact that the rank of the SLP matrix is one, as stated in the following \textit{Corollary}.
	
	\textit{Corollary}: The rank of the optimized precoding matrix ${\bf{P}}$ in \text{P7} is one. 
	
	\textit{Proof}: The precoding matrix can be decomposed into vectors
	\begin{equation}\label{W}
		{\bf{P}} = \left[ {{{\bf{p}}_1},{{\bf{p}}_2}, \cdots ,{{\bf{p}}_{KL}}} \right].
	\end{equation}
	Each ${\bf{p}}_i{s_i} $ is identical based on the work in \cite{21}, then we can further transform the power budget into
	\begin{equation}\label{power}
		\left\| {{\bf{Ps}}} \right\|_2^2 = {\left( {KL} \right)^2}\left\| {{{\bf{p}}_i}{s_i}} \right\|_2^2 = KL\sum\limits_{i = 1}^{KL} {s_i^*{\bf{p}}_i^H{{\bf{p}}_i}{s_i}},
	\end{equation}
	then the problem can be further expressed as
	\begin{equation}\label{P8}
		\begin{array}{l}
			{\rm{\text{P8}}}:\mathop {\max }\limits_{{{\bf{p}}_i},t} {\kern 1pt} {\kern 1pt} {\kern 1pt} {\kern 1pt} t \vspace{1ex} \\ 
			{\kern 20pt}  s.t.{\kern 10pt} {{\bf{F}}_k}{\sum\limits_{i = 1}^{KL} {{{\bf{p}}_i}{s_i}} } = {{\bf{\Lambda }}_k}{{\bf{s}}_k},\forall k \in {\cal K} \vspace{1ex}  \\
			{\kern 42pt}  \left| {\lambda _{k,l}^{\cal J}} \right| - \left( {\lambda _{k,l}^{\cal R} - t} \right)\tan {\theta _t} \le 0, \forall k \in {\cal K},\forall l \in {\cal L} \vspace{1ex} \\
			{\kern 42pt}  \sum\limits_{i = 1}^{KL} {s_i^*{\bf{p}}_i^H{{\bf{p}}_i}{s_i}}  - \dfrac{{{P_T}}}{{KL}} \le 0. \\
		\end{array}
	\end{equation}
	
	In the following, we analyze \text{P8} with Lagrangian and KKT conditions. The Lagrangian of \text{P8} is expressed as\cite{19}
	\begin{equation}\label{L7}
		\begin{array}{l}
			{\cal L}\left( {{{\bf{p}}_i},t,{{\mathord{\buildrel{\lower3pt\hbox{$\scriptscriptstyle\frown$}} 
							\over \mu } }_{k,l}},{{\mathord{\buildrel{\lower3pt\hbox{$\scriptscriptstyle\frown$}} 
							\over \nu } }_{k,l}},{{\mathord{\buildrel{\lower3pt\hbox{$\scriptscriptstyle\frown$}} 
							\over \alpha } }_0}} \right) \vspace{1ex}\\
			=  - t + \sum\limits_{k = 1}^K {\sum\limits_{l = 1}^L {{{\mathord{\buildrel{\lower3pt\hbox{$\scriptscriptstyle\frown$}} 
								\over \mu } }_{k,l}}\left( {{{\bf{f}}_{k,l}}\sum\limits_{i = 1}^{KL} {{{\bf{p}}_i}{s_i}}  - {\lambda _{k,l}}{s_{k,l}}} \right)} }  \vspace{1ex}\\
			{\kern 8pt}+ \sum\limits_{k = 1}^K {\sum\limits_{l = 1}^L {{{\mathord{\buildrel{\lower3pt\hbox{$\scriptscriptstyle\frown$}} 
								\over \nu } }_{k,l}}\left( {\left| {\lambda _{k,l}^{\cal J}} \right| - \left( {\lambda _{k,l}^{\cal R} - t} \right)\tan {\theta _t}} \right)} }  \vspace{1ex}\\
			{\kern 8pt}+ {{\mathord{\buildrel{\lower3pt\hbox{$\scriptscriptstyle\frown$}} 
						\over \alpha } }_0}\left( {\sum\limits_{i = 1}^{KL} {s_i^*{\bf{p}}_i^H{{\bf{p}}_i}{s_i}}  - \dfrac{{{P_T}}}{{KL}}} \right),
		\end{array}
	\end{equation}
	where ${{\bf{f}}_{k,l}}$ is the $l$-th row of ${{\bf{F}}_{k}}$. The ${{{\mathord{\buildrel{\lower3pt\hbox{$\scriptscriptstyle\frown$}}\over \mu } }_{k,l}}}$, ${{{\mathord{\buildrel{\lower3pt\hbox{$\scriptscriptstyle\frown$}}\over \nu } }_{k,l}}} \geq 0$ and ${{{\mathord{\buildrel{\lower3pt\hbox{$\scriptscriptstyle\frown$}}\over \alpha } }_{0}}} \geq 0$ are the dual variables. Based on the Lagrangian in (\ref{L7}), the KKT conditions for optimality can be obtained as 
	\begin{subequations}\label{KKT4}
		\begin{align}
			&\label{e1}\frac{{\partial {\cal L}}}{{\partial {{\bf{p}}_i}}} = \left( {\sum\limits_{k = 1}^K {\sum\limits_{l = 1}^L {{{\mathord{\buildrel{\lower3pt\hbox{$\scriptscriptstyle\frown$}} 
									\over \mu } }_{k,l}}{{\bf{f}}_{k,l}}} } } \right){s_i} + {\mathord{\buildrel{\lower3pt\hbox{$\scriptscriptstyle\frown$}} 
					\over \alpha } _0}{s_i}s_i^*{\bf{p}}_i^H = {\bf{0}},{\kern 1pt} {\kern 1pt} {\kern 1pt} \forall i \in {\cal K}{\cal L}\\  
			&\label{e2}\frac{{\partial {\cal L}}}{{\partial t}} = \sum\limits_{k = 1}^K {\sum\limits_{l = 1}^L {{{\mathord{\buildrel{\lower3pt\hbox{$\scriptscriptstyle\frown$}} 
								\over \nu } }_{k,l}}\tan {\theta _t}} }  - 1 = 0\\  
			&\label{e3} {{{\bf{f}}_{k,l}}\sum\limits_{i = 1}^{KL} {{{\bf{p}}_i}{s_i}}  - {\lambda _{k,l}}{s_{k,l}}} =0,\forall k \in {\cal K},\forall l \in {\cal L}\\
			&\label{e4}{{{\mathord{\buildrel{\lower3pt\hbox{$\scriptscriptstyle\frown$}} 
							\over \nu } }_{k,l}}\left( {\left| {\lambda _{k,l}^{\cal J}} \right| - \left( {\lambda _{k,l}^{\cal R} - t} \right)\tan {\theta _t}} \right)}=0,\forall k \in {\cal K},\forall l \in {\cal L}\\
			&\label{e5}{{\mathord{\buildrel{\lower3pt\hbox{$\scriptscriptstyle\frown$}} 
						\over \alpha } }_0}\left( {\sum\limits_{i = 1}^{KL} {s_i^*{\bf{p}}_i^H{{\bf{p}}_i}{s_i}}  - \dfrac{{{P_T}}}{{KL}}} \right)=0,\forall k \in {\cal K},\forall l \in {\cal L}.
		\end{align}
	\end{subequations}
	Based on (\ref{e1}), we can obtain that ${{{\mathord{\buildrel{\lower3pt\hbox{$\scriptscriptstyle\frown$}}\over \alpha } }_{0}}} \ne 0$, and with the fact that ${{{\mathord{\buildrel{\lower3pt\hbox{$\scriptscriptstyle\frown$}}\over \alpha } }_{0}}} \geq 0$, we can further have ${{{\mathord{\buildrel{\lower3pt\hbox{$\scriptscriptstyle\frown$}}\over \alpha } }_{0}}} > 0$. Then, the ${\bf{w}}^H_i$ in (\ref{e1}) can be expressed as
	\begin{equation}\label{wiH}
		\begin{array}{l}
			{\bf{p}}_i^H =  - \left( {\sum\limits_{k = 1}^K {\sum\limits_{l = 1}^L {\dfrac{{{{\mathord{\buildrel{\lower3pt\hbox{$\scriptscriptstyle\frown$}} 
											\over \mu } }_{k,l}}}}{{{{\mathord{\buildrel{\lower3pt\hbox{$\scriptscriptstyle\frown$}} 
											\over \alpha } }_0}}}{{\bf{f}}_{k,l}}} } } \right) \cdot \dfrac{1}{{s_i^*}},\forall i \in {\cal K}{\cal L}.
		\end{array}
	\end{equation}
	By introducing
	\begin{equation}\label{lambdakl}
		\begin{array}{l}
			{\lambda _{k,l}} =- \dfrac{{\mathord{\buildrel{\lower3pt\hbox{$\scriptscriptstyle\frown$}} 
						\over \mu } _{k,l}^H}}{{{{\mathord{\buildrel{\lower3pt\hbox{$\scriptscriptstyle\frown$}} 
								\over \alpha } }_0}}},\forall k \in {\cal K},\forall l \in {\cal L},
		\end{array}
	\end{equation}
	the expression of ${\bf{p}}_i$ is obtained as
	\begin{equation}\label{wi}
		\begin{array}{l}
			{\bf{p}}_i =   \left( {\sum\limits_{k = 1}^K {\sum\limits_{l = 1}^L {{\lambda _{k,l}}{{\bf{f}}_{k,l}^H}} } } \right) \cdot \dfrac{1}{{s_i}},\forall i \in {\cal K}{\cal L}.
		\end{array}
	\end{equation}
	Based on (\ref{wi}), we can further obtain
	\begin{equation}\label{wisi}
		\begin{array}{l}
			{{\bf{p}}_i}{s_i} = \sum\limits_{k = 1}^K {\sum\limits_{l = 1}^L {{\lambda _{k,l}}{\bf{f}}_{k,l}^H} } , \forall i \in {\cal K}{\cal L},
		\end{array}
	\end{equation}
	which means that ${{\bf{p}}_i}{s_i}$ is constant for any $i$. Additionally, this observation confirms that the precoding vector for one symbol is a phase-rotated version of the another precoding vector within a channel coherence interval, which leads to $rank({\bf{P}}) = 1$. 
	
	\qquad \qquad\qquad\qquad\qquad\qquad\qquad\qquad\qquad\qquad\qquad\quad\ \ $\qedsymbol$
	
	To further demonstrate that the rank-one precoding matrix leads to infeasible solutions for the receive combining matrix, we provide \textbf{Lemma 1} and its accompanying \textit{proof}:
	
	{\bf{Lemma 1:}} Assuming that ${\bf{A}}{_{m \times n}}$ and ${\bf{B}}{_{n \times s}}$ , then we have $rank({\bf{AB}})\leq min\{rank({\bf{A}}),rank({\bf{B}})\}$.
	
	\textit{Proof:} First, the auxiliary matrix ${\bf{C}} $ is introduced and expressed as
		\begin{equation}\label{C}
			{\bf{C}} = {\bf{AB}} = {\left[ {{{\bf{c}}_1},{{\bf{c}}_2}, \cdots ,{{\bf{c}}_m}} \right]^T},
		\end{equation}
		then we have
		\begin{equation}\label{AB}
			\begin{array}{l}
				{\bf{AB}} = \left[ {\begin{array}{*{20}{c}}
						{{a_{11}}}&{{a_{12}}}& \cdots &{{a_{1n}}}\\
						{{a_{21}}}&{{a_{22}}}& \cdots &{{a_{2n}}}\\
						\vdots & \vdots & \ddots & \vdots \\
						{{a_{m1}}}&{{a_{m2}}}& \cdots &{{a_{mn}}}
				\end{array}} \right]\left[ {\begin{array}{*{20}{c}}
						{{{\bf{b}}_1}}\\
						{{{\bf{b}}_2}}\\
						\vdots \\
						{{{\bf{b}}_n}}
				\end{array}} \right] \vspace{1ex}\\
				{\kern 17pt}= \left[ {\begin{array}{*{20}{c}}
						{{a_{11}}{{\bf{b}}_1} + {a_{12}}{{\bf{b}}_2} +  \cdots {a_{1n}}{{\bf{b}}_n}}\\
						{{a_{21}}{{\bf{b}}_1} + {a_{22}}{{\bf{b}}_2} +  \cdots {a_{2n}}{{\bf{b}}_n}}\\
						\vdots \\
						{{a_{m1}}{{\bf{b}}_1} + {a_{m2}}{{\bf{b}}_2} +  \cdots {a_{mn}}{{\bf{b}}_n}}
				\end{array}} \right] \vspace{1ex}\\
				{\kern 17pt}= \left[ {\begin{array}{*{20}{c}}
						{{{\bf{c}}_1}}\\
						{{{\bf{c}}_2}}\\
						\vdots \\
						{{{\bf{c}}_m}}
				\end{array}} \right],
			\end{array}
		\end{equation}
		where ${\bf{B}} = {\left[ {{{\bf{b}}_1},{{\bf{b}}_2}, \cdots ,{{\bf{b}}_n}} \right]^T}$. According to \cite{34}, we know that for vector sets $\left\{ {{{\bm{\alpha }}_1},{{\bm{\alpha }}_2}, \cdots ,{{\bm{\alpha }}_u}} \right\}$ and $\left\{ {{{\bm{\beta }}_1},{{\bm{\beta }}_2}, \cdots ,{{\bm{\beta }}_t}} \right\}$, if any of ${{\bm{\beta }}_i}\left( {i = 1,2, \cdots ,t} \right)$ can be linearly expressed by $\left\{ {{{\bm{\alpha }}_1},{{\bm{\alpha }}_2}, \cdots ,{{\bm{\alpha }}_u}} \right\}$, then it represents 
		\begin{equation}\label{rank}
			rank\left\{ {{{\bm{\beta }}_1},{{\bm{\beta }}_1}, \cdots ,{{\bm{\beta }}_t}} \right\} \le rank\left\{ {{{\bm{\alpha }}_1},{{\bm{\alpha }}_1}, \cdots ,{{\bm{\alpha }}_u}} \right\}.
		\end{equation}
		It's shown that anyone of ${\bf{c}}_i$ can be linearly expressed by row vectors of $\bf{B}$ in (\ref{AB}), and based on the (\ref{rank}), we have
		\begin{equation}\label{ABB}
			rank({\bf{AB}}) \le rank({\bf{B}}).
		\end{equation}
		Similarly, we can proof that $rank({\bf{AB}}) \le rank({\bf{A}})$. Therefore, we have 
		\begin{equation}\label{lemma1}
			rank({\bf{AB}})\leq min\{rank({\bf{A}}),rank({\bf{B}})\}.
		\end{equation}
		\qquad \qquad\qquad\qquad\qquad\qquad\qquad\qquad\qquad\qquad\qquad\qquad\ \ $\qedsymbol$

	
	Based on \textbf{Lemma 1}, we can establish that
	\begin{equation}\label{rank1}
		rank({\bf{G}}k^H{\bf{R}}\eta^{-1}{{\bf{G}}_k})= 1,
	\end{equation}
	where ${{\bf{G}}_k} = {{\bf{H}}_k}{{\bf{P}}_k}$. When deriving the receive combining matrix ${\bf{W}}_k$, it is necessary to compute the inverse of ${\bf{G}}k^H{\bf{R}}\eta^{-1}{{\bf{G}}_k}$. However, the rank-one property of ${\bf{G}}k^H{\bf{R}}\eta^{-1}{{\bf{G}}_k}$ makes the inverse infeasible, thereby making the calculation of ${\bf{W}}_k$ infeasible as well. 
	
	\subsection{Regularized IRC Receiver}\label{transprecoding}
	In order to solve the rank-one IRC matrix problem, the RIRC receiver is firstly proposed in our paper. The technique often used to ``regularize'' an inverse is to add a multiple of the identity matrix before inverting. For example, instead of calculating ${\bf{W}}_k$ using (\ref{IRC}), we use 
	\begin{equation}\label{pk1}
		{{\bf{W}}_k} = {\left( {{\bf{G}}_k^H{\bf{R}}_\eta ^{ - 1}{{\bf{G}}_k} + \gamma {\bf{I}}} \right)^{ - 1}}{\bf{G}}_k^H{\bf{R}}_\eta ^{ - 1},
	\end{equation}
	where $\gamma$ is the regularized factor. Then, the AO algorithm is used to update ${\bf{W}}_k$ and ${\bf{P}}$ iteratively, which is shown in \textbf{Algorithm \ref{al2}}.
	\begin{algorithm}[!h]
		\caption{Proposed AO algorithm for solving problem (\ref{P8})}
		\label{al2}
		\begin{algorithmic}
			\State ${\bf Input:}$ ${\bf{H}}$, ${\bf{s}}$, and $\kappa$.\vspace{0.5ex}
			\State ${\bf Output:}$ ${\bf{P}}$, ${\bf{W}}_k$\vspace{0.5ex}
			\State ${\bf Initialization:}$  Set the variable ${\bf{P}}^{(1)}$ with BD algorithm, $t^{(1)}=0$, $ \kappa=10{^{-5}}$, and iteration index $ \textit{n}=1 $\vspace{0.5ex}
			\While {$\delta > \kappa $} \vspace{0.5ex}
			\State Obtain ${{\bf{W}}_{k}^{(n+1)}} $  via (\ref{pk1})\vspace{0.5ex}
			\State Obtain ${{\bf{\hat u}}}$ via (\ref{P3}) \vspace{0.5ex}
			\State Obtain ${{\bf{P}}^{(n+1)}}$ and $t^{(n+1)}$ via (\ref{closed-form W}) and (\ref{P2}) \vspace{0.5ex}
			\State $\delta  = \left| t^{(n+1)}-t^{(n)} \right|$
			\State $n=n+1$
			\EndWhile
			\State ${\bf P}={\bf P}^{\left( {n} \right)}$, ${{\bf{W}}_k}={\bf{W}}_k^{\left( {n} \right)}$.
		\end{algorithmic}
	\end{algorithm}
	
	For the given \{${{\bf{P}}^{(n)}} $, ${{\bf{W}}_k^{(n)}} $\}, the solutions \{${{\bf{P}}^{(n+1)}} $, ${{\bf{W}}_k^{(n+1)}} $\} obtained in the ($ n$+1) iteration by using (\ref{closed-form W}) and (\ref{pk1}). In each iteration, the solution is locally optimal and the objective function value $t^{(n+1)}$ is non-decreasing. When the difference value between the two iterations is less than convergence threshold $\kappa$, the algorithm is terminated. Furthermore, the convergence and computational complexity analysis of the proposed \textbf{Algorithm \ref{al2}} is in Section \ref{CCCA}.
	
	\section{Convergence and Computational Complexity Analysis} \label{CCCA}
	In this section, we analyze the convergence and computational costs of the proposed two AO algorithms. Firstly, the convergence is guaranteed by analyzing the algorithms' monotonicity and boundedness, then the computational complexity analysis is given for both of the proposed algorithms and \textbf{Algorithm \ref{al2}} enjoys lower computational complexity.
	\subsection{Convergence Analysis}
	The AO method is used in both of our proposed algorithms, and the convergence of the proposed AO algorithms can be guaranteed for following two reasons: 1) Within each iteration of the proposed AO algorithms, the value of the objective function monotonically increases. This property ensures progress towards a better solution in each iteration, leading to convergence\cite{24}. 2) The objective function value has an upper bound for both of the proposed problems, which is determined by the non-strict CI constraints. This upper bound acts as a convergence criterion, ensuring that the optimization process does not diverge and remains within a feasible range\cite{25}. The convergence of the AO algorithms is further supported by the numerical results presented in Section \ref{Simulation}. These results validate that the proposed AO algorithms indeed converge and effectively solve the symbol level transmit precoding and receive combining design problem for MU-MIMO system.
	\subsection{Computational Complexity Analysis}\label{CCA}
	The total complexity of the proposed \textbf{Algorithm \ref{al1}} can be divided into two parts: the optimization of receive combining matrix ${\bf{W}}_k$ and the calculation of precoding matrix ${\bf{P}}$. On the one hand, the CVX is used to obtain the dual variables ${\bf{\tilde u}}$ and ${{\bf{\hat u}}}$, the complexity cost are all ${\mathcal{O}}({(2KL)}^{3.5})$\cite{33}, since there are the same number of variables in the two convex problem. On the other hand, the main overhead of the computation of the two closed-form solutions is the matrix inverse operation in $\bf{P}$, which costs about ${\mathcal{O}}({K^{3}L^{3}})$. Therefore, by denoting $N_{iter1}$ as the number iterations, we can obtain the total computational complexity of  \textbf{Algorithm \ref{al1}} as $N_{iter1}(2{\mathcal{O}}({(2KL)}^{3.5})+{\mathcal{O}}({K^{3}L^{3}}))$. 
	
	Similar as the complexity analysis of \textbf{Algorithm \ref{al1}}, the total complexity of the proposed \textbf{Algorithm \ref{al2}} consists of the calculation of the closed-form solutions and convex QP problem, but only one QP problem needs to be solve by using CVX, which reduces the computational complexity. We define $N_{iter2}$ as the number iterations of \textbf{Algorithm \ref{al2}}, it can be observed that $N_{iter2}<N_{iter1}$ in Section \ref{Simulation}. Then, we can obtain the total computational complexity of  \textbf{Algorithm \ref{al2}} as $N_{iter2}({\mathcal{O}}({(2KL)}^{3.5})+{\mathcal{O}}({K^{3}L^{3}}))$. Obviously, the lack of the calculation of one dual variable simplifies the computation process, which makes \textbf{Algorithm \ref{al2}} enjoy lower computational complexity.
	
	\section{Simulation Results}\label{Simulation}
	In this section, the numerical results of the proposed two schemes are presented and compared with the traditional transmit BD precoding with IRC receiver approach\cite{25, 35} based on the Monte Carlo simulations. In each slot, the transmit power budget is set as ${P_T}=1$, the transmit SNR is expressed as $\rho=1/ {\sigma^2}$, and each entry in ${\bf{H}}$ follows a standard complex Gaussian distribution as ${\bf{H}}_{m,n} \in \mathcal{CN}(0,1) $. Furthermore, the regularized factor is chosen to be 1, because larger factor value will lead to more performance degradation caused by the calculation error. To clarify, the following abbreviations are used throughout this section:

	\begin{enumerate}
		\item `BD+IRC': the transmitter uses the traditional BD algorithm to preprocess the transmit signal and the IRC receiver is used to decode the received information signal, which is represented by the blue line;
		\item `Joint Design': joint symbol level transmit precoding and receive combining design based on \textbf{Algorithm \ref{al1}}, which is represented by the purple line;
		\item `Iterative SLP+RIRC': symbol level transmit precoding with RIRC receiver based on \textbf{Algorithm \ref{al2}}, which is represented by the yellow line;
		\item `SLP+RIRC': in order to reduce the computational complexity, after initializing  the precoding matrix ${\bf{P}}^{(1)}$ with BD method and IRC matrix ${\bf{W}}_k^{(1)}$, the ${\bf{P}}$ and ${\bf{W}}_k$ are directly calculated by using (\ref{closed-form W}) and (\ref{pk1}), which is represented by the red line.
	\end{enumerate}

	\begin{figure}[!t]
		\centering
		\includegraphics[width=3.5in]{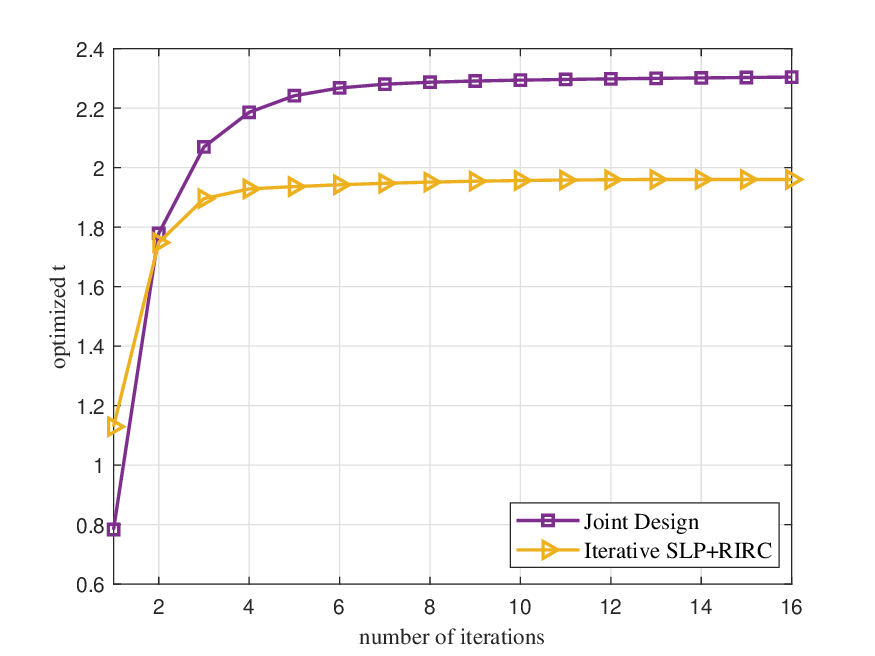}
		\caption{Convergence for proposed algorithms.}
		\label{convergence}
	\end{figure}
	\begin{figure}[!t]
		\centering
		\includegraphics[width=3.5in]{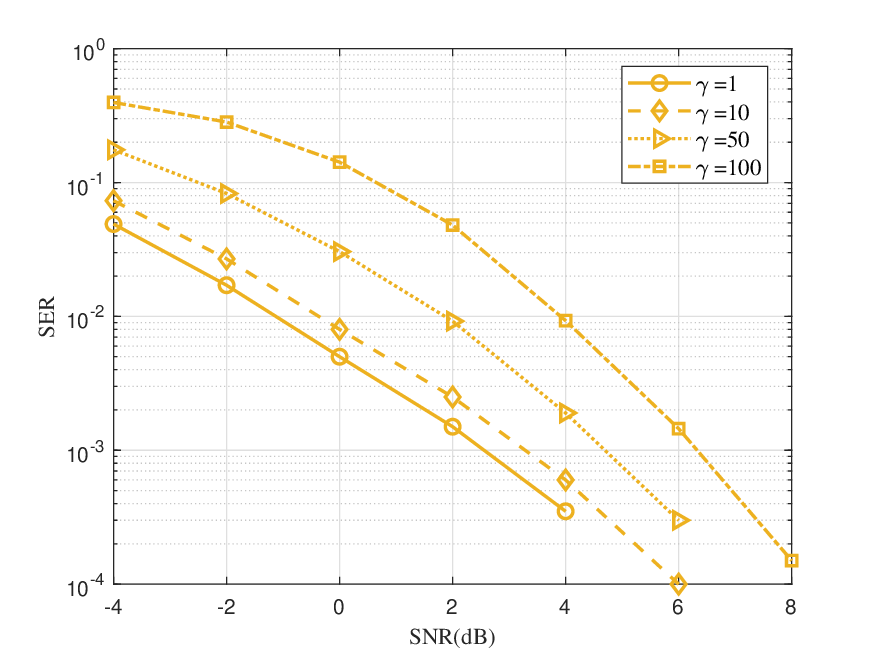}
		\caption{SER v.s. transmit SNR with different regularized factor $\gamma$.}
		\label{gamma}
	\end{figure}
	
	Fig. \ref{convergence} illustrates the convergence of the two proposed algorithms. It is evident that both algorithms converge only within several iterations. However, it is observed that the `Joint Design' approach requires more iterations to converge compared to the `Iterative SLP+RIRC' scheme. This can be attributed to the fact that the `Iterative SLP+RIRC' scheme eliminates the calculation of one of the dual variables, allowing for direct derivation of the receive combing matrix. This finding is consistent with the computational complexity analysis presented in Section \ref{CCA}. Furthermore, it is worth noting that the `Joint Design' approach converges to a higher value than the `Iterative SLP+RIRC' scheme. This difference indicates that the practical `Iterative SLP+RIRC' scheme experiences a performance loss due to the lack of transmit symbol information. This observation also depicts the trade-off between computational complexity and performance that exists between the two proposed approaches.
	
	Fig. \ref{gamma} demonstrates the symbol error rate (SER) performance of the proposed `Iterative SLP+RIRC' scheme with different values of the regularized factor $\gamma$. This factor is introduced to address the unsolvable problem arising from the rank-one precoding matrix. It's obvious that the SER performance deteriorates as the value of the regularized factor increases. This performance degradation can be attributed to the fact that larger regularized factor values will result in increased calculation errors, thereby compromising the decoding performance. 
	\begin{figure}[!t]
		\centering
		\includegraphics[width=3.5in]{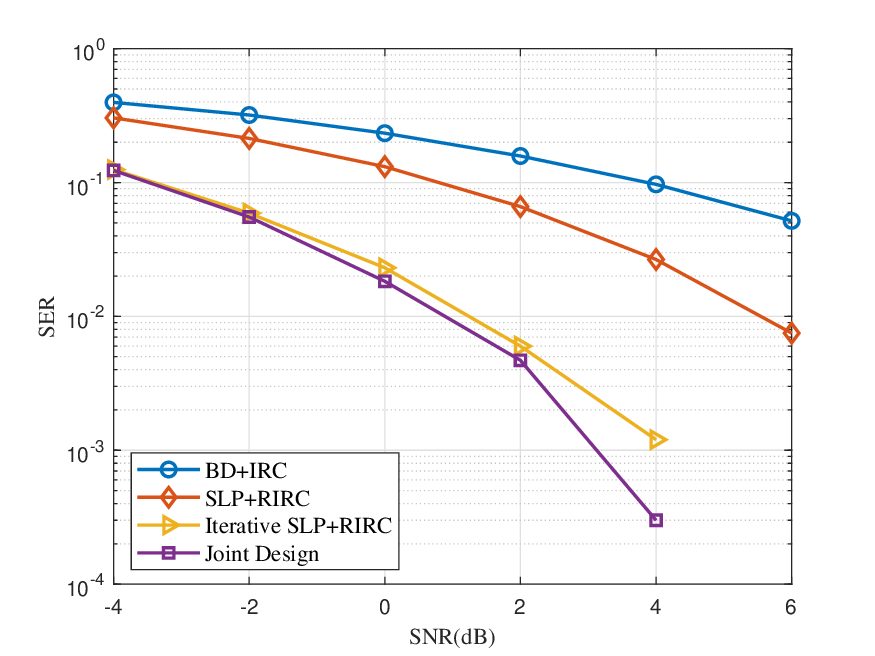}
		\caption{SER v.s. transmit SNR, QPSK, $N_T=8$ and $N_R=K=L=2$.}
		\label{QPSK8222}
	\end{figure}
	\begin{figure}[!t]
		\centering
		\includegraphics[width=3.5in]{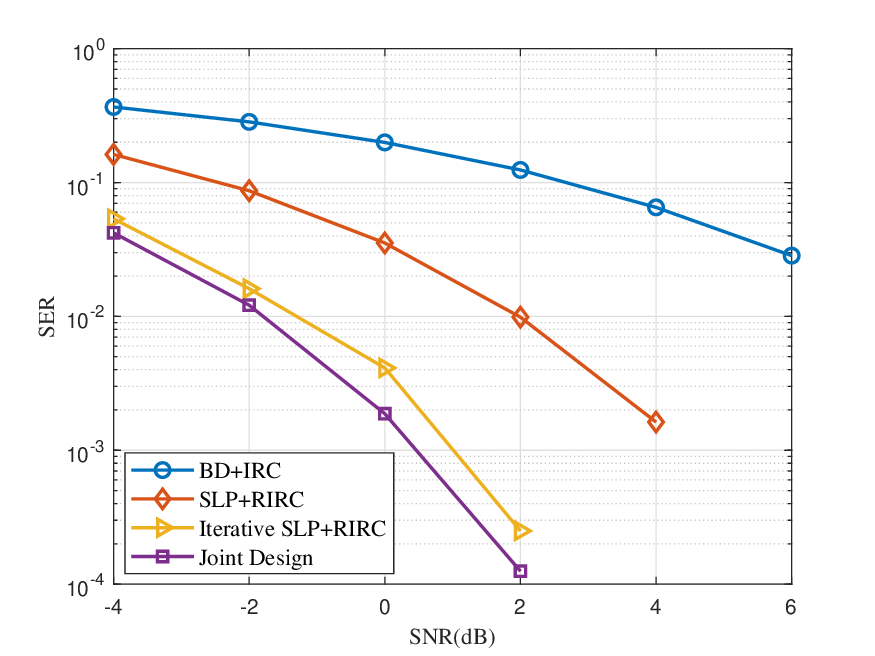}
		\caption{SER v.s. transmit SNR, QPSK, $N_T=8$, $N_R=4$ and $K=L=2$.}
		\label{QPSK8242}
	\end{figure}
	
	In Fig. \ref{QPSK8222} and Fig. \ref{QPSK8242}, we compare the SER performance of different schemes employing QPSK modulation. In both figures, $N_T=8, K=L=2$, while $N_R$ is either $2$ or $4$. It's evident that the proposed SLP schemes achieve an improved performance over the `BD+IRC' scheme for the use of symbol-by-symbol optimization. Moreover, the performance improvement is more significant in the case of `Joint Design' scheme. Additionally, the more practical `Iterative SLP+RIRC' scheme only has an acceptable performance loss compared with the `Joint Design' scheme, and the `SLP+RIRC' scheme enjoys lower computational complexity. The difference between the Fig. \ref{QPSK8222} and Fig. \ref{QPSK8242}  lies in the number of receive antennas. In Fig. \ref{QPSK8222}, the number of receive antennas equals the number of data streams, whereas in Fig. \ref{QPSK8242}, there are more receive antennas than data streams.  When receivers are equipped with more antennas than the data streams, additional degrees of freedom can be utilized to enhance communication performance. Consequently, we observe that Fig. \ref{QPSK8242} achieves a lower SER compared to Fig. \ref{QPSK8222} under the same SNR conditions. Furthermore, it is evident that higher SER gains can be achieved at the high SNR regime, which is consistent with the findings in reference \cite{19}. 
	
%
	
\begin{figure}[!t]
	\centering
	\includegraphics[width=3.5in]{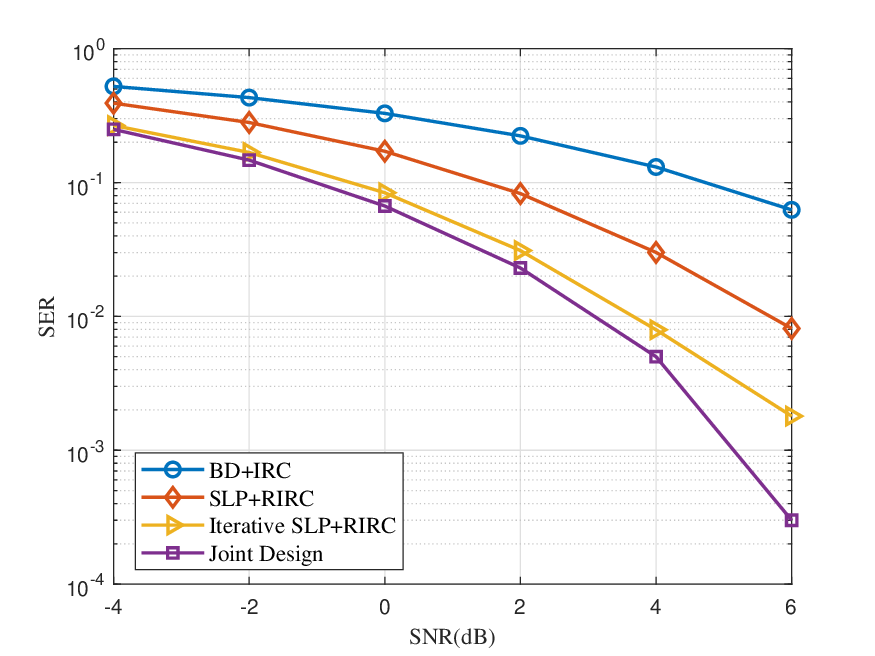}
	\caption{SER v.s. transmit SNR, 8PSK, $N_T=16$ and $N_R=K=L=2$.}
	\label{8PSK16222}
\end{figure}
\begin{figure}[!t]
	\centering
	\includegraphics[width=3.5in]{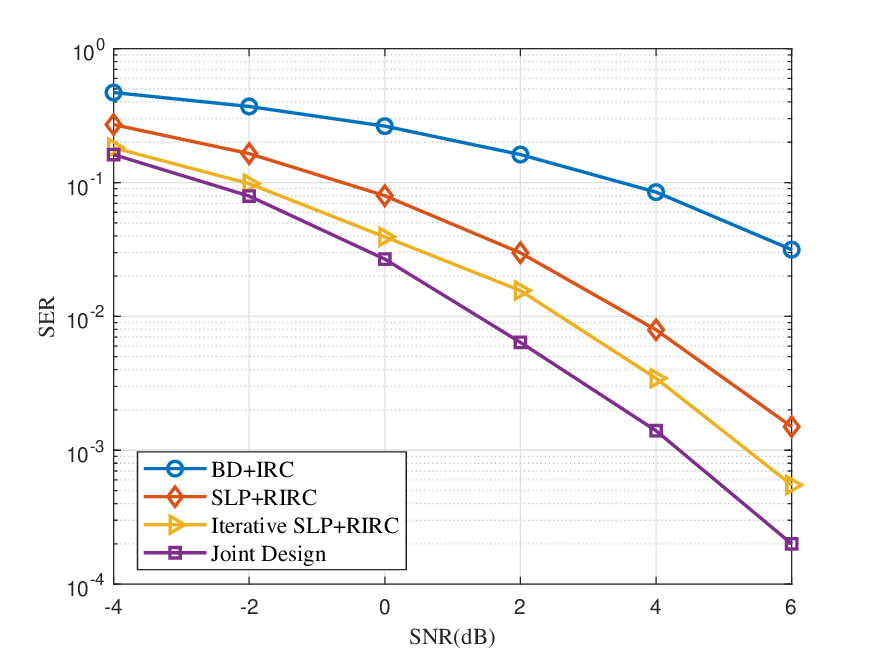}
	\caption{SER v.s. transmit SNR, 8PSK, $N_T=16$, $ N_R=4$ and $K=L=2$.}
	\label{8PSK16242}
\end{figure}	
	
	Fig. \ref{8PSK16222} and Fig. \ref{8PSK16242} also depict the SER performance of different schemes with 8PSK modulation. In the two figures, $N_T=16, K=L=2$, while $N_R$ is either $2$ or $4$, where $ N_T > KN_R $. The similar SER trends can be observed that SER decreases with the increase of the transmit SNR. Similarly, the proposed SLP schemes outperform the traditional `BD+IRC' scheme. Furthermore, it is shown that higher SER gains can be achieved at the high SNR regime, especially in the Fig. \ref{8PSK16242}, where the number of data streams is less than the number of the receive antennas. Additionally, we observe that Fig. \ref{8PSK16222} achieves a lower SER compared to Fig. \ref{8PSK16242} under the same SNR conditions.
	
	
	\begin{figure}[!t]
		\centering
		\includegraphics[width=3.5in]{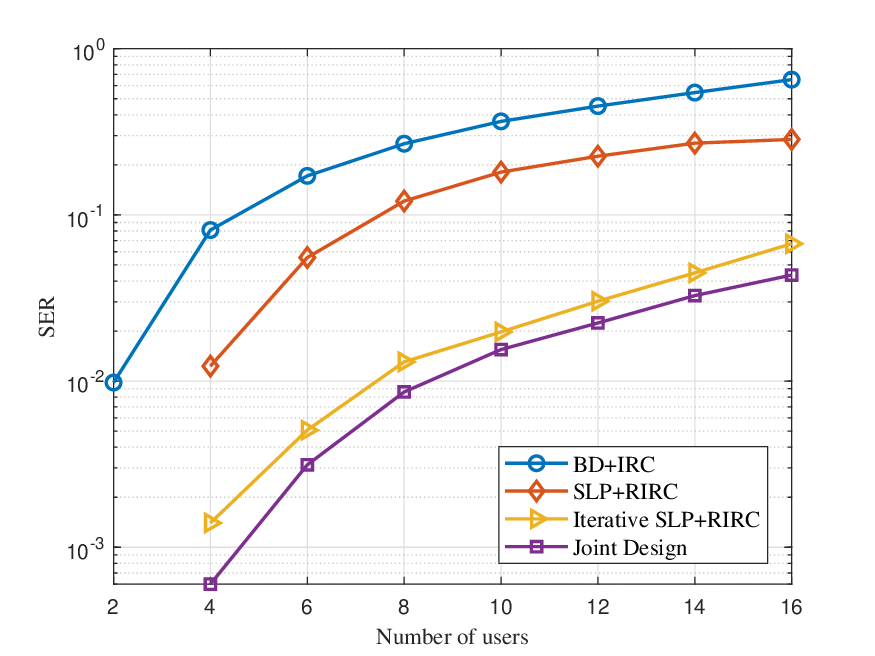}
		\caption{SER v.s. Number of users, QPSK, $N_T=32$ and $N_R=L=2$.}
		\label{SERvsU}
	\end{figure}
	\begin{figure}[!t]
		\centering
		\includegraphics[width=3.5in]{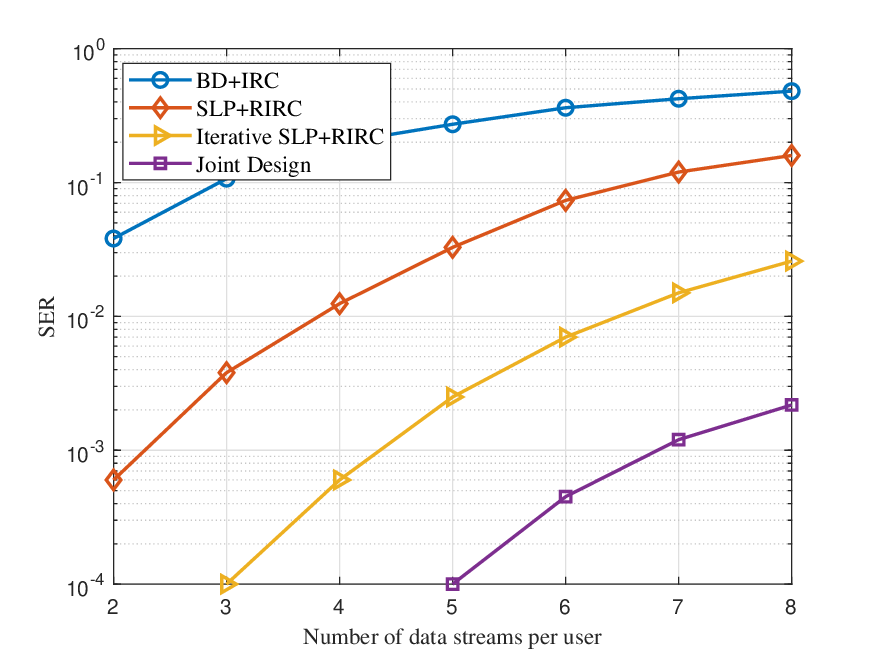}
		\caption{SER v.s. Number of data streams per user, QPSK, $N_T=16$, $N_R=8$ and $K=2$.}
		\label{SERvsDS}
	\end{figure}
	\begin{figure}[!t]
		\centering
		\includegraphics[width=3.5in]{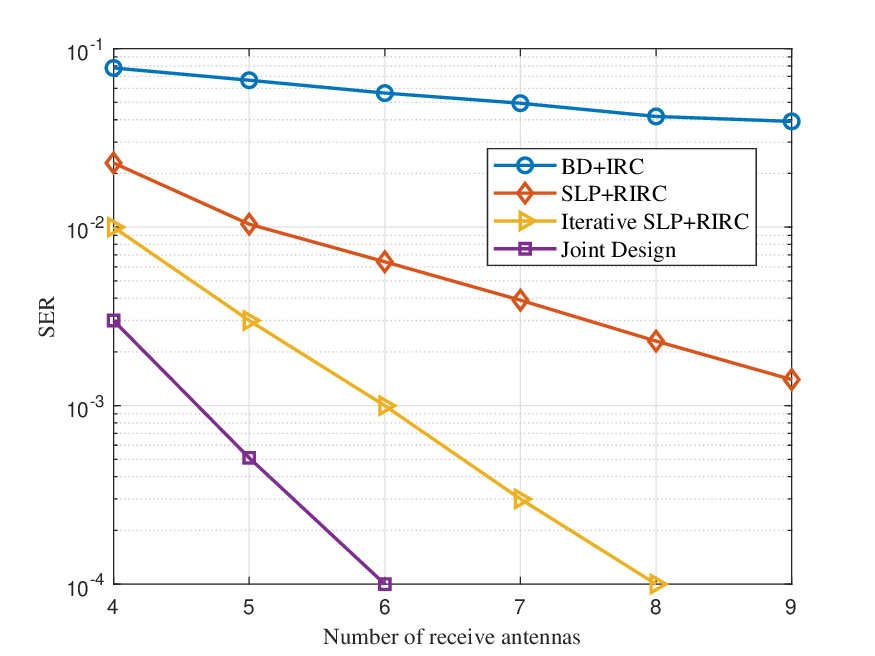}
		\caption{SER v.s. Number of receive antennas, QPSK, $N_T=36$, $K=2$ and $L=4$.}
		\label{SERvsRA}
	\end{figure}

	Fig. \ref{SERvsU} depicts the SER performance with respect to the number of users employing QPSK modulation in different schemes. It's evident that the SER increases as the number of users increases while keeping other parameters constant. This is attributed to the fact that the increasing number of users results in fewer design freedoms available, until $N_T=KN_R$. At this point, the one-to-one mapping between transmit antennas and receive antennas imposes performance limitations. Consistent with the aforementioned results, the `Joint Design' approach demonstrates the best SER performance among the considered schemes. Moreover, the practical `Iterative SLP+RIRC' scheme exhibits only a slight performance loss compared to the `Joint Design' approach. 
	
	In Fig. \ref{SERvsDS}, it can be observed that the SER increases as the number of data streams grows. This can be attributed to similar reasons as in Fig. \ref{SERvsU}, the increase of the data streams decreases the design freedoms until $N_R=L$, leading to one-to-one mapping between the receive antennas and data streams. Despite this, the proposed schemes still outperform the traditional BD-based algorithm.
	
	Fig. \ref{SERvsRA} demonstrates a notable trend where the SER decreases as the number of receive antennas increases. This is because of that as the number of receive antennas grows, more design freedoms are available to more effective design of the receive combining matrix, which confirms the positive impact of additional receive antennas on the system's ability to mitigate errors and enhance overall communication performance.

	\section{Conclusion}\label{Conclusion}
	This paper proposes SLP designs for MU-MIMO communication systems, where the BS transmits multiple data streams to multi-antenna users. By analyzing the Lagrangian and KKT conditions, the optimal structures of the joint transmit precoding and the receive combining are obtained and simpler optimization problems are derived by formulating the dual problem. Additionally, a more practical approach called the `Iterative SLP+RIRC' scheme is proposed to overcome the problem where the receive combining matrix is related to the transmit signal in the joint design scheme. Numerical results reveal that the practical `Iterative SLP+RIRC' method exhibits only a marginal communication performance loss when compared to the joint transmit precoding and receive combining design. However, both of the `Iterative SLP+RIRC' and joint design approaches offer significant performance gains over conventional BD-based methods.


\begin{thebibliography}{00}
		
		\bibitem{1} 
		T. M. Berhane, W. X. Meng, L. Chen, G. D. Jobir and C. Li, “SLNR-Based Precoding for Single Cell Full-Duplex MU-MIMO Systems,” \textit{IEEE Transactions on Vehicular Technology.}, vol. 66, no. 9, pp. 7877-7887, 2017.
		
		
		\bibitem{2} 
		A. Z. Piracha, H. Farid, K. Shahzad and M. Zeeshan, ``A Low Complexity Signal-to-Total Variance Precoding Scheme for Downlink Multi-Stream MU-MIMO Systems,'' \textit{2022 3rd International Conference on Electrical Engineering and Informatics (ICon EEI).}, Pekanbaru, Indonesia, pp. 53-58, 2022.
		
		\bibitem{3} 
		M. Sadek, A. Tarighat and A. H. Sayed, ``A Leakage-Based Precoding Scheme for Downlink Multi-User MIMO Channels,'' \textit{IEEE Transactions on Wireless Communications.}, vol. 6, no. 5, pp. 1711-1721, 2007.
		
		\bibitem{M1} 
		L. Li \textit{et al}. ``mmWave communications for 5G: implementation challenges and advances,'' \textit{Science China-Information Sciences.}, 021301, 2018.
		
		\bibitem{M2} 
		P. -D. Arapoglou, K. Liolis, M. Bertinelli, A. Panagopoulos, P. Cottis and R. De Gaudenzi, ``MIMO over Satellite: A Review,'' \textit{IEEE Communications Surveys \& Tutorials.}, vol. 13, no. 1, pp. 27-51, 2011.
		
		\bibitem{M3} 
		J. Fan, Z. Xu and G. Y. Li, ``Performance Analysis of MU-MIMO in Downlink Cellular Networks,'' \textit{IEEE Communications Letters.}, vol. 19, no. 2, pp. 223-226, 2015.
		
		\bibitem{M4} 
		M. Ghosh, “A Comparison of Normalizations for ZF Precoded MU-MIMO Systems in Multipath Fading Channels,” \textit{IEEE Wireless Communications Letters.}, vol. 2, no. 5, pp. 515-518, 2013.
		
		\bibitem{4} 
		M. Costa, “Writing on dirty paper,” \textit{IEEE Transactions on Information Theory.}, vol. IT-29,no. 3, pp. 439–441, 1983.
		
		\bibitem{5} 
		L. Sun and M. Lei, “Quantized CSI-Based Tomlinson-Harashima Precoding in Multiuser MIMO Systems,” \textit{IEEE Transactions on Wireless Communications.}, vol. 12, no. 3, pp. 1118-1126, 2013.
		
		\bibitem{6} 
		B. M. Hochwald, C. B. Peel and A. L. Swindlehurst, “A vector-perturbation technique for near-capacity multiantenna multiuser communication-part II: perturbation,” \textit{IEEE Transactions on Communications.}, vol. 53, no. 3, pp. 537-544, 2005.
		
		\bibitem{7} 
		A. Wiesel, Y. C. Eldar and S. Shamai, “Zero-Forcing Precoding and Generalized Inverses,” \textit{IEEE Transactions on Signal Processing.}, vol. 56, no. 9, pp. 4409-4418, 2008.
		
		\bibitem{8} 
		C. B. Peel, B. M. Hochwald and A. L. Swindlehurst, “A vector-perturbation technique for near-capacity multiantenna multiuser communication-part I: channel inversion and regularization,” \textit{IEEE Transactions on Communications.}, vol. 53, no. 1, pp. 195-202, 2005.
		
		\bibitem{252}
		Q. H. Spencer, A. L. Swindlehurst and M. Haardt, “Zero-forcing methods for downlink spatial multiplexing in multiuser MIMO channels,” \textit{IEEE Transactions on Signal Processing.}, vol. 52, no. 2, pp. 461-471, 2004.
		
		\bibitem{26}
		K. Zu, R. C. de Lamare and M. Haardt, “Generalized Design of Low-Complexity Block Diagonalization Type Precoding Algorithms for Multiuser MIMO Systems,” \textit{IEEE Transactions on Communications.}, vol. 61, no. 10, pp. 4232-4242, 2013.
		
		\bibitem{27}
		H. Sung, S. R. Lee and I. Lee, “Generalized Channel Inversion Methods for Multiuser MIMO Systems,” \textit{IEEE Transactions on Communications.}, vol. 57, no. 11, pp. 3489-3499, 2009.
		
		\bibitem{9} 
		Kai-Kit Wong, R. D. Murch and K. B. Letaief, “Performance enhancement of multiuser MIMO wireless communication systems,” \textit{IEEE Transactions on Communications.}, vol. 50, no. 12, pp. 1960-1970, 2002.
		
		\bibitem{10} 
		M. F. Hanif, L. N. Tran, A. Tölli and M. Juntti, “Computationally Efficient Robust Beamforming for SINR Balancing in Multicell Downlink With Applications to Large Antenna Array Systems,” \textit{IEEE Transactions on Communications.}, vol. 62, no. 6, pp. 1908-1920, 2014.
		
		\bibitem{11} 
		F. Wang, X. Wang and Y. Zhu, “Transmit beamforming for multiuser downlink with per-antenna power constraints,” \textit{2014 IEEE International Conference on Communications (ICC).}, Sydney, NSW, Australia, pp. 4692-4697, 2014.
		
		\bibitem{12}
		P. Cheng, M. Tao and W. Zhang, ``A New SLNR-Based Linear Precoding for Downlink Multi-User Multi-Stream MIMO Systems,'' \textit{IEEE Communications Letters.}, vol. 14, no. 11, pp. 1008-1010, 2010.
		
		\bibitem{131} 
		A. Li et al. "A Tutorial on Interference Exploitation via Symbol-Level Precoding: Overview, State-of-the-Art and Future Directions," \textit{IEEE Communications Surveys \& Tutorials.}, vol. 22, no. 2, pp. 796-839, 2020.
		
		\bibitem{13} 
		A. Li \textit{et al}. “A Tutorial on Interference Exploitation via Symbol-Level Precoding: Overview, State-of-the-Art and Future Directions,” \textit{IEEE Communications Surveys \& Tutorials.}, vol. 22, no. 2, pp. 796-839, 2020.
		
		\bibitem{14} 
		A. Li, L. Song, B. Vucetic and Y. Li, “Interference Exploitation Precoding for Reconfigurable Intelligent Surface Aided Multi-User Communications With Direct Links,” \textit{IEEE Wireless Communications Letters.}, vol. 9, no. 11, pp. 1937-1941, 2020.
		
		\bibitem{15} 
		A. Li, F. Liu, C. Masouros, Y. Li and B. Vucetic, “Interference Exploitation 1-Bit Massive MIMO Precoding: A Partial Branch-and-Bound Solution With Near-Optimal Performance,” \textit{IEEE Transactions on Wireless Communications.}, vol. 19, no. 5, pp. 3474-3489, 2020.
		
		\bibitem{16} 
		Y. Qin, X. Liao, A. Li and C. Masouros, “Low-Complexity PAPR Minimization for Symbol Level Precoded Multi-User MISO-OFDM System,” \textit{IEEE Communications Letters.}, vol. 26, no. 2, pp. 409-413, 2022.
		
		\bibitem{17} 
		Y. Fan, A. Li, X. Liao and V. C. M. Leung, “Secure Interference Exploitation Precoding in MISO Wiretap Channel: Destructive Region Redefinition With Efficient Solutions,” \textit{IEEE Transactions on Information Forensics and Security.}, vol. 16, pp. 402-417, 2021.
		
		\bibitem{18} 
		Q. Xu, P. Ren and A. L. Swindlehurst, “Rethinking Secure Precoding via Interference Exploitation: A Smart Eavesdropper Perspective,” \textit{IEEE Transactions on Information Forensics and Security.}, vol. 16, pp. 585-600, 2021.
		
		\bibitem{21} 
		C. Masouros and G. Zheng, “Exploiting Known Interference as Green Signal Power for Downlink Beamforming Optimization,” \textit{IEEE Transactions on Signal Processing.}, vol. 63, no. 14, pp. 3628-3640, 2015.
		
		\bibitem{19} 
		A. Li and C. Masouros, “Interference Exploitation Precoding Made Practical: Optimal Closed-Form Solutions for PSK Modulations,” \textit{IEEE Transactions on Wireless Communications.}, vol. 17, no. 11, pp. 7661-7676, 2018.
		
		\bibitem{20} 
		A. Li, C. Masouros, B. Vucetic, Y. Li and A. L. Swindlehurst, “Interference Exploitation Precoding for Multi-Level Modulations: Closed-Form Solutions,” \textit{IEEE Transactions on Communications.}, vol. 69, no. 1, pp. 291-308, 2021.
		
		\bibitem{22} 
		A. Li, C. Shen, X. Liao, C. Masouros and A. Lee Swindlehurst, “Practical Interference Exploitation Precoding without Symbol-by-Symbol Optimization: A Block-Level Approach,” \textit{IEEE Transactions on Wireless Communications.}, Early Access, pp. 1-1, 2022.
		
		\bibitem{23}
		Z. Xiao, R. Liu, M. Li, Y. Liu and Q. Liu, “Low-Complexity Designs of Symbol-Level Precoding for MU-MISO Systems,” \textit{IEEE Transactions on Communications.}, vol. 70, no. 7, pp. 4624-4639, 2022.
		
		\bibitem{24}
		Q. Li, M. Hong, H. T. Wai, Y. F. Liu, W. K. Ma and Z. Q. Luo, ``Transmit Solutions for MIMO Wiretap Channels using Alternating Optimization,'' \textit{IEEE Journal on Selected Areas in Communications}, vol. 31, no. 9, pp. 1714-1727, 2013.
		
		\bibitem{25}
		A. A. Salem, M. H. Ismail and A. S. Ibrahim, ``Active Reconfigurable Intelligent Surface-Assisted MISO Integrated Sensing and Communication Systems for Secure Operation,'' \textit{IEEE Transactions on Vehicular Technology}, vol. 72, no. 4, pp. 4919-4931, 2023.
		
		\bibitem{31}
		S. Cai, T. H. Chang and H. Zhu, “Joint Symbol Level Precoding and Receive Beamforming for Multiuser MIMO Downlink,” \textit{2019 IEEE 20th International Workshop on Signal Processing Advances in Wireless Communications (SPAWC).}, Cannes, France, pp. 1-5, 2019.
		
		\bibitem{32}
		S. Cai, H. Zhu, C. Shen and T. H. Chang, “Joint Symbol Level Precoding and Receive Beamforming Optimization for Multiuser MIMO Downlink,” \textit{IEEE Transactions on Signal Processing.}, vol. 70, pp. 6185-6199, 2022.
		
		\bibitem{35}
		Z. Bai et al.,``On the Equivalence of MMSE and IRC Receiver in MU-MIMO Systems," \textit{IEEE Communications Letters}, vol. 15, no. 12, pp. 1288-1290,2011.
		
		\bibitem{33}
		L. Vandenberghe and S. Boyd, Convex Optimization. Cambridge, U.K.: Cambridge Univ. Press, 2004.
		
		\bibitem{34}
		X. D. Zhang, Matrix analysis and application. Tsinghua, C.H.N.: Tsinghua Univ. Press, 2013.
		
	\end{thebibliography}
\end{document}